\documentclass{aa}
\usepackage{tablefootnote}
\usepackage{graphicx, xcolor}
\usepackage[space]{grffile}
\usepackage{latexsym}
\usepackage{float}
\usepackage{booktabs}
\usepackage{amsfonts,amsmath,amssymb}
\usepackage{url}
\usepackage{hyperref}
\usepackage{multirow}
\usepackage{xcolor}
\usepackage[normalem]{ulem}
\usepackage{txfonts}

\title{Generative Bayesian modelling of radio antenna beams}
\subtitle{An application to the MeerKAT radio telescope}

\author{Morteza Pashapour-Ahmadabadi$^{1}$\thanks{\email{morteza.pasha@physik.uni-bielefeld.de}}, 
Dominik J.\ Schwarz$^{1}$,
Mattieu S. de Villiers$^2$,
Aritra Basu$^{3,4,5}$, 
Vincent Eberle$^6$,
Torsten En{\ss}lin$^{3,6,7}$,
Philipp Frank$^{6,8}$, 
Richard Fuchs$^{9}$,
Hans-Rainer Kl\"ockner$^{5}$,
Anna Nelles$^{10,11}$
}

\institute{$^{1}$ Fakult\"at f\"ur Physik, Universit\"at Bielefeld, Postfach 100131, 33501 Bielefeld, Germany\\
$^2$ South African Radio Astronomy Observatory, Liesbeek House, River Park, Mowbray, 7700, Cape Town, South Africa\\
$^3$ Deutsches Zentrum f\"ur Astrophysik, Postplatz 1, 02826 G\"orlitz, Germany\\
$^4$ Th\"uringer Landessternwarte, Sternwarte 5, 07778 Tautenburg, Germany\\
$^5$ Max-Planck-Institut f\"ur Radioastronomie, Auf dem H\"ugel 69, 53121 Bonn, Germany\\
$^6$ Max-Planck-Institut f\"ur Astrophysik, Karl-Schwarzschild-Str. 1, 85748 Garching, Germany\\
$^7$ Ludwig-Maximilians-Universit\"at M\"unchen, Geschwister-Scholl-Platz 1, 80539 Munich, Germany\\
$^8$ Kavli Institute for Particle Astrophysics \& Cosmology (KIPAC), Stanford University, CA 94305 Stanford, USA\\
$^{9}$ Technical University of Munich, TUM School of Natural Sciences, Boltzmannstr. 2, 85748 Garching, Germany\\
$^{10}$ Erlangen Centre for Astroparticle Physics (ECAP), Friedrich-Alexander-University Erlangen-N\"urnberg, 91058 Erlangen, Germany\\
$^{11}$ Deutsches Elektronen-Synchrotron DESY, Platanenallee 6, 15738 Zeuthen, Germany\\
}

\authorrunning{Pashapour-Ahmadabadi et al.}

\date{Received XXX ; accepted XXX}
    
\begin{document} 

  \abstract
{Accurate primary beam models are essential for high-fidelity calibration and imaging in radio astronomy. Traditional approaches either lack the ability to represent realistic antenna patterns (analytical models) or could become computationally prohibitive to interpolate for wide-bandwidth observations (electromagnetic simulations, direct holographic measurements).}
{We develop a generative beam model that
represents both the spatial structure and the frequency variations in the primary beam, while remaining computationally tractable for integration into the calibration and imaging pipelines. We aim to achieve superior accuracy compared to existing parametric models while reducing data storage and computational requirements compared to data-based models.}
{We adopted a non-parametric Bayesian framework grounded in information field theory, implemented using the NIFTy library. The model represents the beam as a base pattern subject to frequency-dependent geometric transformations (scaling and shifting). 
The Model performance was validated through posterior predictive checks ($\chi^2/\mathrm{dof} \approx 1$) and compared it against traditional beam models (Gaussian, Airy disc, JimBeam, Eidos, electromagnetic simulation). We applied the new method to holographic data of the array-averaged MeerKAT Stokes I beam in the L band.}
{The model achieves 400:1 data compression, while preserving statistical consistency with holographic measurements. It outperforms existing approaches by more than an order of magnitude in mean squared deviation across the MeerKAT L band. The Zernike decomposition of the base beam, combined with frequency-dependent geometric transformations, enables computationally efficient beam reconstruction at an arbitrary spatial resolution.
}
{We demonstrate that generative Bayesian modelling provides an accurate, compact, and physically interpretable representation of radio telescope primary beams. The 150 kB per-antenna footprint enables memory slim integration into calibration and imaging software at negligible computational cost. The approach scales favourably to large arrays and wide bandwidths. The methodology can be readily adapted to other radio telescopes, offering a general pathway towards improved calibration precision for current and future radio surveys.}
    
\maketitle

\section{Introduction}

The accurate characterisation of an antenna’s beam is essential for observational studies in radio astronomy. An incorrect beam model can significantly degrade image quality by introducing systematic flux scale distortions that are particularly severe for off-axis sources far from the phase centre. 
Precise characterisation of the beam sidelobes is especially important in observations that include bright sources outside the target field, as inaccurate sidelobe modelling can lead to spurious flux density contamination and imaging artefacts. 
Such inaccuracies introduce artefacts that cannot be mitigated by additional temporal or spectral integration.

A wide range of beam models has been developed for different types of antennas, ranging from purely theoretical formulations to data-driven approaches that incorporate measurement-based corrections (\citealt{silver1949}; \citealt{Ruze1966}; \citealt{Harrington1968}; \citealt{ScottRyle1977}; \citealt{RahmatSamii1980APA}; \citealt{Olver1994}; \citealt{Hamaker1996}; \citealt{CondonRansom2016}; \citealt{Perley_2017}; \citealt{Iheanetu2019}; \citealt{Asad2021}; \citealt{Kern2025}; \citealt{Wilensky2025}). In recent years, with the advent of increasingly complex radio arrays operating under diverse observing conditions, the need for precise beam modelling has grown substantially. Capturing the detailed behaviour of the antenna beam not only improves calibration efficiency but also enhances image fidelity, thereby enabling more accurate scientific measurements and discoveries.

The objective of this study is to develop a beam model based on a non-parametric Bayesian framework grounded in information field theory (IFT; \citealt{Ensslin2009}), capable of capturing both the spatial and spectral structure of the beam in a compact and physically meaningful way. This theory provides a theoretical foundation for extending Bayesian inference to continuous fields. Building on this foundation, we employ NIFTy (numerical information field theory; \citealt{Niftyre2013}; \citealt{arras2019nifty}; \citealt{niftyre}), a computational library specifically designed for Bayesian inference in high-dimensional and function-space problems. NIFTy offers tools for estimating statistical properties of complex posterior distributions from noisy and incomplete data and has demonstrated scalability to problems involving half a billion parameters (\citealt{niftyre}). By minimising the number of degrees of freedom, the model aims to represent only the physically relevant components and their variations. 

In this work, we study the generative Bayesian modelling of the antenna beam of a typical modern radio telescope operating in the mid radio-frequency range (1–10 GHz). We detail the GAMBIT (generative antenna model from Bayesian inference technique) framework, first sketched in \citet{Pasha2026SPIE}. 
These telescopes typically consist of a parabolic primary reflector and can have receiving elements in an on-axis primary focus—as the Australian Square Kilometre Array Pathfinder (ASKAP) antennas \citep{ASKAP2021} or the planned Deep Synoptic Array (DSA; \citealt{DSA2019}) do — or use an offset Gregorian design, as the MeerKAT \citep{Meerkat2016} and SKA-Mid \citep{SKAmid2015} telescopes do. While ASKAP antennas have the additional complexity of phased array feeds (PAFs), and since DSA, MeerKAT Extension and SKA-Mid are still under construction, we choose MeerKAT for this study due to the availability of high-quality holographic beam measurements across its full operational bandwidth.
Holographic measurements characterise the beam by observing a strong, known calibrator source, while scanning the antenna across a grid of pointing offsets, directly mapping the complex beam response across the field of view. Unlike electromagnetic (EM) simulations—which rely on idealised models of antenna reflector and feed geometry, holographic measurements provide a direct, high signal-to-noise, frequency-resolved characterisation of the actual as-built antenna response, capturing real-world imperfections and frequency-dependent effects. We anticipate that lessons learned for MeerKAT antenna beams will be directly applicable to the future characterisation of SKA-Mid antenna beams. 

MeerKAT is a radio interferometer comprising 64 offset Gregorian dishes, each with an effective diameter of D=13.5\,m, distributed over baselines of up to 8\,km in the Northern Cape Province of South Africa \citep{Meerkat2016}. The MeerKAT antenna beam has previously been modelled using several approaches. Among these are full-wave EM simulations \citep{deValliers2019,Asad2021,deVilliers2021OMT}, employing detailed three-dimensional representations of the antenna structure. Although based on the exact physics of Maxwell's equations, these simulations necessarily rely on idealised as-designed geometries: modelling every structural detail is computationally intractable. As a result, EM simulations do not fully capture the frequency-dependent pointing behaviour 
\citep{deVilliers2022}, which arises from real-world effects absent from the idealised model. A second efficient approach for beam modelling involves characteristic basis function patterns, which decompose complex beam structures into a small number of representative basis functions. The full beam pattern can then be reconstructed by summing over the dominant `eigenbeams,' resulting in a set of simpler components that capture the essential beam characteristics. This technique enables fast and scalable beam predictions, making it especially suitable for large interferometric arrays like MeerKAT (\citealt{Maskant2012}; \citealt{Mutonkole2016}; \citealt{deValliers2018}). An alternative approach characterises the beam using a unique set of Zernike polynomials (\citealt{Zernike1934}; \citealt{BornWolf1999}), which provide compact mathematical descriptions of the beam's spatial structure and spectral behaviour. 

Existing approaches to primary beam modelling lie on a spectrum that balances flexibility against computational cost. Using holographic measurement data from MeerKAT antennas \citep{deVilliers2023}, we develop a compact, generative beam model that accurately represents the key characteristics of the measured antenna response across the L band (856–1712 MHz), while remaining computationally efficient enough to enable rapid beam evaluation through linear combinations. We compare our NIFTy-based beam model with several existing approaches, ranging from analytical parametric formulations to data-driven methods.

Sect.~\ref{sec:data-characteristics} details the characteristic features of the holographically measured MeerKAT antenna beam in the L band. In Sect.~\ref{sec:beam-model} we introduce several parametric models commonly used to fit antenna beams in radio astronomy. Sect.~\ref{sec:nifty-model} is dedicated to the generative Bayesian beam modelling framework and its application to MeerKAT data. We discuss the physical properties of the inferred beam model and compare its performance to existing approaches in Sect.~\ref{sec:meerkat-beam}, and conclude in Sect.~\ref{sec:conclusion}. In Appendix~\ref{App:ShiftParameterComparison} we present a centre-of-mass analysis of shift parameters across multiple antennas, and in Appendix~\ref{App:NoiseAnalysis} we examine the noise estimation methodology used in the Bayesian inference. Appendix~\ref{app:ZernikeBaseBeam} examines the dominant Zernike polynomial modes of the inferred base beam.

\section{Data characteristics}
\label{sec:data-characteristics}
To establish a generative Bayesian beam model and to compare it with alternative beam models, we use holographic beam measurements of the MeerKAT antenna in the L band, covering the frequency range from 856 to 1712 MHz across 1024 frequency channels, as provided by \citet{deVilliers2023}. The data were downloaded from the MeerKAT Holographic Measurement Archive.\footnote{\url{https://doi.org/10.48479/wdb0-h061}} The scan extent parameters for these measurements were chosen to encompass approximately two sidelobes at the lowest frequencies and up to five sidelobes at the highest frequencies within the L band. This ensures adequate coverage of the beam's spatial structure across the full spectral range. Although the full Jones matrix is available for each frequency channel, in this study we focus on modelling the Stokes I primary beam. Modelling the full Jones matrix is considerably more intricate because it represents a complex-valued tensor field. In contrast, the Stokes I beam is a real-valued scalar field, making it more tractable for initial modelling eﬀorts. It is computed as
\begin{equation}
    I = \frac{1}{2}\left(|E_{HH}|^2 + |E_{HV}|^2 + |E_{VH}|^2 + |E_{VV}|^2\right),
\end{equation}
where $E_{HH}$, $E_{HV}$, $E_{VH}$, and $E_{VV}$ are the four complex Jones matrix elements. Modelling the full polarisation response is deferred to future work.

\begin{figure}
    \centering
    \includegraphics[width=\linewidth]{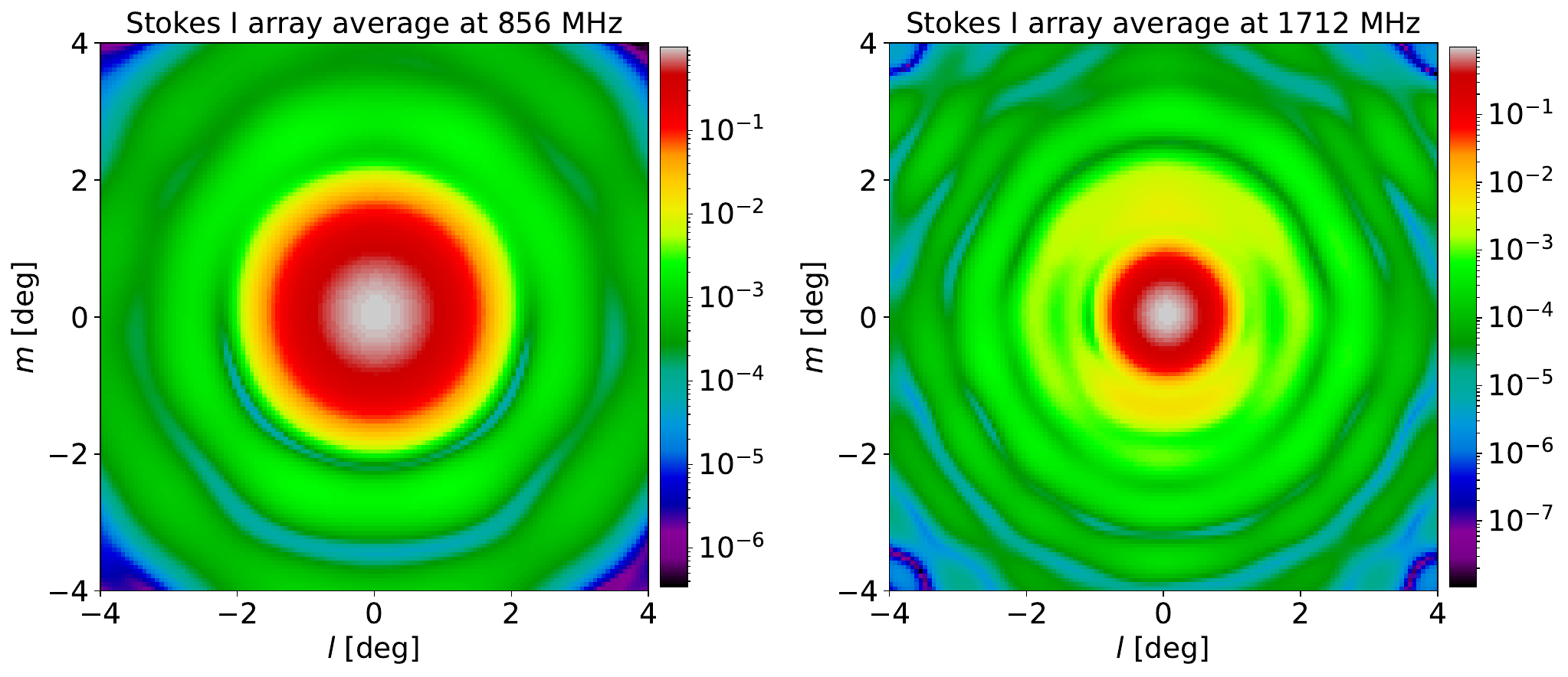}
\caption{Measured Stokes I array-averaged beam from holographic observations of the MeerKAT antennas at the lower (856~MHz, left) and upper (1712~MHz, right) edges of the L band. The beam is shown on a logarithmic colour scale (normalised intensity) spanning approximately seven orders of magnitude.}
    \label{StokesI_2Channels}
\end{figure}

\begin{figure}
    \centering
    \includegraphics[width=\linewidth]{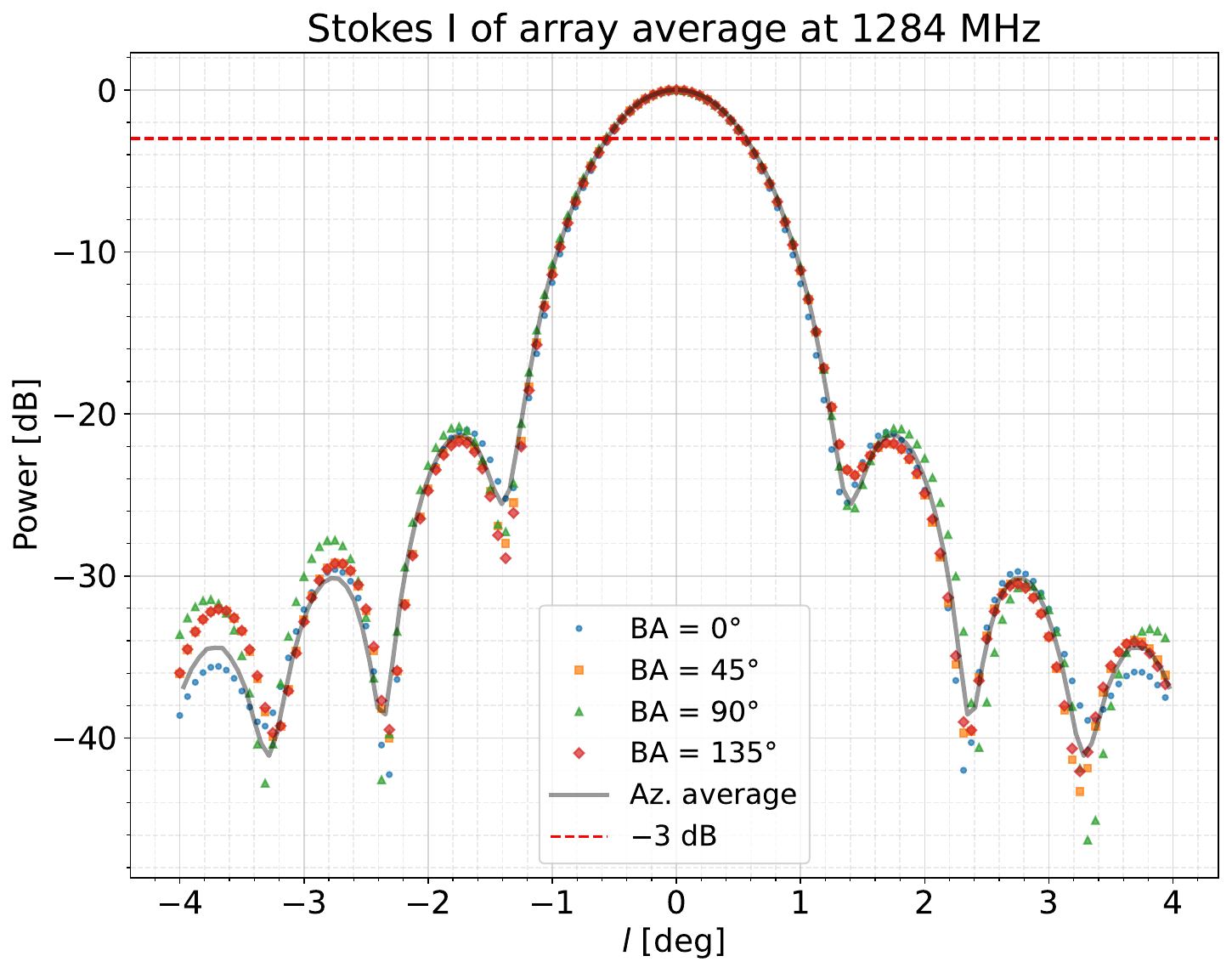}
    \caption{%
        Measured beam pattern for the array average at 1284\,MHz showing power (dB) 
        versus angular offset at four beam angles (BA\,=\,$0\degr$, $45\degr$, $90\degr$, $135\degr$) and its azimuthal average (solid black). The dashed red line marks the $-3$\,dB level (FWHM\,$\approx 1\degr$). The main lobe and multiple sidelobes 
        extending to $\pm4\degr$ are clearly visible, and show asymmetry in the detailed beam structure of the holographic measurements.}
        
    \label{fig:FWHM_1284MHz}
\end{figure}

Fig.~\ref{StokesI_2Channels} shows the MeerKAT array-averaged Stokes I beam at two representative frequency channels. The array-averaged beam is constructed by averaging holographic measurements across all 64 antennas, providing a single characteristic beam pattern representative of the typical MeerKAT response. The coordinate system is aligned with the antenna's feed reference frame: the $m$-axis is oriented in the elevation (vertical) direction, while the $l$-axis is oriented in the azimuth (horizontal) direction. Each beam map spans $8\degr \times  8\degr$, sampled on a uniform $128 \times 128$ pixel-grid corresponding to a pixel size $0.0625\degr$ (3.75\,arcmin).

Fig.~\ref{fig:FWHM_1284MHz} shows the measured beam pattern for the MeerKAT array average at 1284\,MHz (central L band frequency). The plot displays beam power (dB) versus angular offset for four beam angles BA\,=\,$0\degr$, $45\degr$, $90\degr$, $135\degr$ and the azimuthal average (solid black). Beam angle refers to the azimuthal angle of the beam in the feed coordinate system and is counted anticlockwise from the positive $l$-direction. Comparing multiple beam angles reveals any directional asymmetries in the beam structure. The azimuthal average represents the beam power averaged over all angles at each radial distance, providing a direction-independent profile.

\begin{figure}
    \centering
    \includegraphics[width=\linewidth]{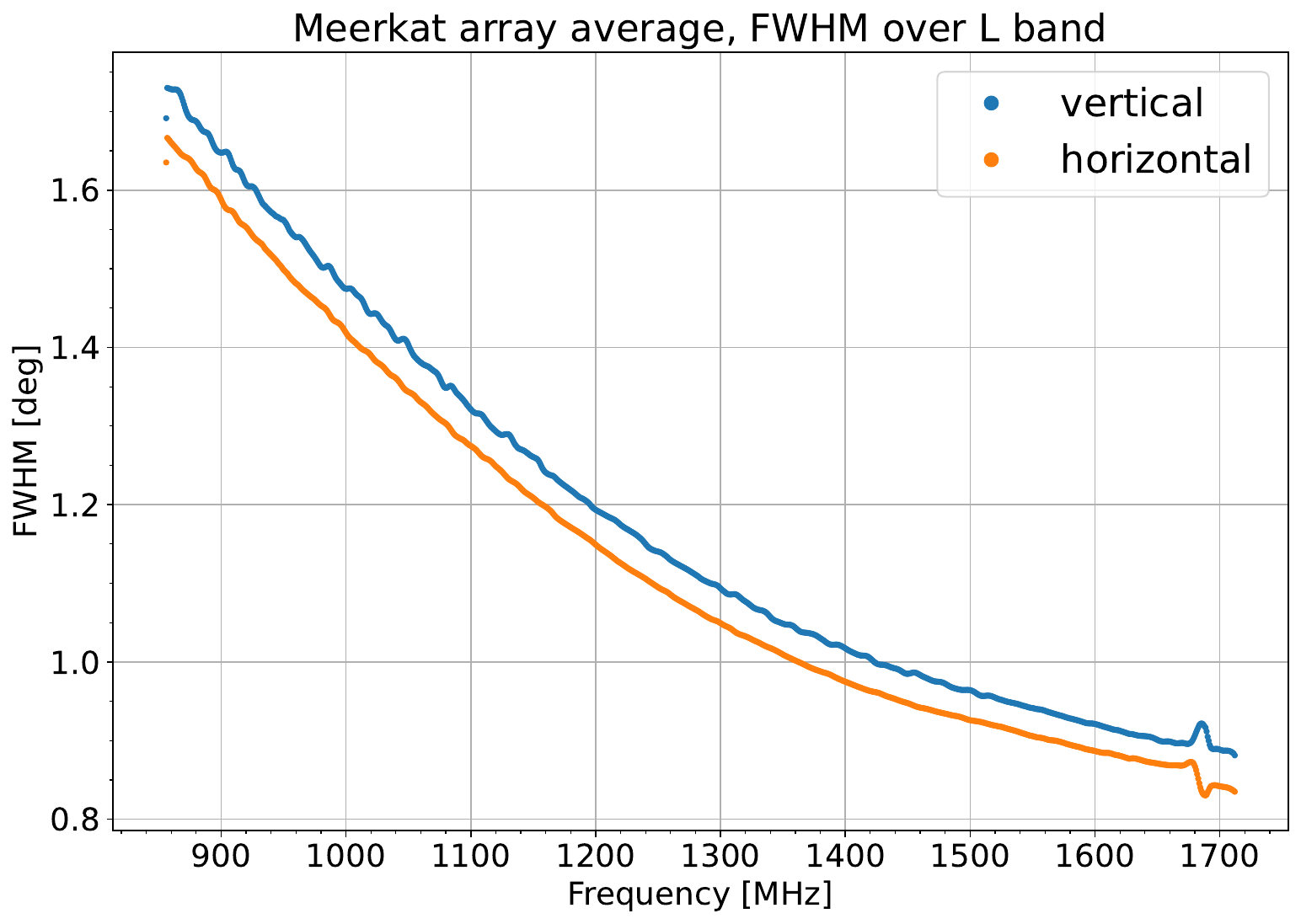}
    \caption{Measured FWHM for array average across the L band along vertical (blue) and horizontal (orange) directions. The beam narrows from $\sim 1.65\degr$ to $\sim 0.85\degr$ following the expected $\lambda/D$ dependence over the bandwidth.t The sharp anomaly near the upper end of the band due to OMT effects.}
    \label{fig:FWHM_over_LBand}
\end{figure}

The dashed red line marks the half-power level, corresponding to a FWHM of ${\sim}1\degr$. The main lobe and several sidelobe rings are clearly resolved out to $\pm4\degr$, illustrating the angular dynamic range afforded by holographic measurements. Fig.~\ref{fig:FWHM_over_LBand} illustrates the FWHM and its variations across the L band. The Stokes I primary beamwidth is generally expected to scale with wavelength $\lambda$, i.e.\ FWHM $\propto \lambda/D$ across most of the band in accordance with diffraction scaling. Across all frequencies $f$, the beam exhibits a noticeable elongation along the vertical axis, where the beamwidth is consistently larger than along the horizontal direction. This asymmetry arises from the antenna’s main reflector geometry, which is wider in the horizontal direction when projected onto the aperture plane. Towards the upper end of the L band, the beamwidth exhibits non-monotonic behaviour, expanding and contracting with frequency. This is attributed to the excitation of the TM11 mode within the waveguide of the orthomode transducer (OMT) that encloses the feed dipoles \citep{deVilliers2023}. Three main characteristics can be observed in the decreasing trend of the FWHM, as shown in Fig.~\ref{fig:FWHM_over_LBand}. First, the overall beamwidth decreases with increasing frequency, approximately following a $1/f$ scaling. However, accurately capturing this frequency dependence requires a more complex model, as the relationship is not linear with frequency due to instrumental and structural factors. Secondly, periodic fluctuations in the FWHM are evident, which have been attributed to the presence of a standing wave between the primary and secondary reflectors of the antenna \citep{Iheanetu2019}. This standing wave induces a frequency-dependent ripple across the entire beam pattern. Thirdly, there appears to be a gradual attenuation of the ripple amplitude at higher frequencies. This effect may be related to reduced reflection efficiency or increased absorption at higher frequencies within the feed or reflector system \citep{Iheanetu2019}. 

\begin{figure}
    \centering
    \includegraphics[width=\linewidth]{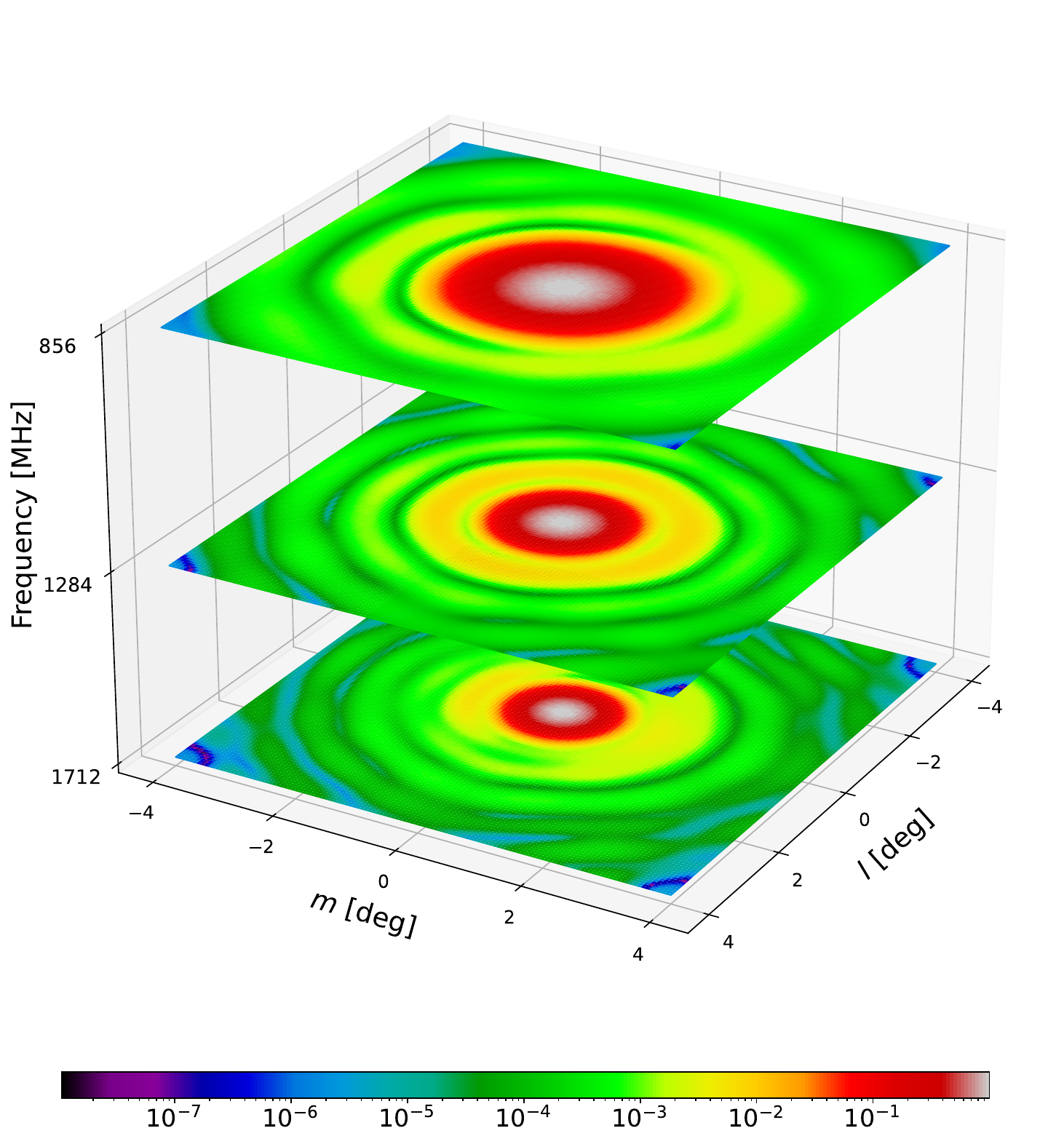}
    \caption{Measured beam patterns for MeerKAT array average (Stokes I) at 900, 1284, and 1700 MHz, showing characteristic beam narrowing with frequency. Spatial extent is $\pm4\degr$ with normalised intensity. These holographic measurements form the input data for the Bayesian reconstruction.}
    \label{fig:3D_StokesI}
\end{figure}

Each holographic measurement forms a three-dimensional dataset, consisting of a stack of two-dimensional spatial beam maps across the full range of observed frequencies, as illustrated in Fig.~\ref{fig:3D_StokesI}. The resulting data cube exhibits a characteristic cone-like shape, embodying the systematic decrease in the FWHM at higher frequencies. The objective of our modelling framework is to compress this high-dimensional dataset---which spans hundreds of frequency channels and tens of thousands of spatial pixels per channel---into a compact, physically meaningful representation that captures both the spatial structure and spectral variations in the beam. This enables rapid beam reconstruction at any frequency with negligible computational cost, while preserving the underlying physical content of the measurements. This model is intended to serve as new, data-driven beam that can be incorporated into the direction-dependent (DD) calibration process, thereby improving the fidelity of calibration in the presence of beam-related systematics. 

\section{Parametric beam models}
\label{sec:beam-model}
We benchmark our model against four established approaches: the Gaussian and Airy disc analytical models, the empirically calibrated JimBeam model, and the Eidos Zernike-based decomposition representing current best practices for modelling MeerKAT holographic beam images.

\subsection{Two-dimensional Gaussian beam model}

The 2D Gaussian beam model approximates the antenna response by fitting an elliptical Gaussian function to the measured holographic data. This simple analytical form has historically served as a practical approximation for the beam pattern of parabolic antennas. For classical radio astronomy applications using large single dishes or small interferometric arrays, the Gaussian approximation is adequate within the FWHM. This was sufficient for instruments with large dish diameters $D$ and correspondingly small fields of view, where scientific targets were typically confined to the central beam region and sidelobe contributions were either negligible or could be empirically corrected. 

The model is defined by five parameters and has the following functional form
\begin{equation}
F_{\mathrm{G}}(l, m) = 
\exp\left\{-\left[\frac{(l - l_0)^2}{2\sigma_{l}^2} + \frac{(m - m_0)^2}{2\sigma_{m}^2}\right]\right\},
\label{eq:gaussian_basic}
\end{equation}
where the coordinates $(l, m)$ represent angular offsets from the pointing centre, respectively, 
$(l_0, m_0)$ is the location of the beam maximum (mean of the Gaussian), and $\sigma_{l}$, $\sigma_{m}$, are the standard deviations along the $l$ and $m$ directions (before rotation), respectively.

When the beam is rotated by a beam angle (BA), the coordinates entering the Gaussian are transformed about the beam centre $(l_0, m_0)$ according to
\begin{align}
l'-l_0 &= (l - l_0)\cos\mathrm{BA} - (m - m_0)\sin\mathrm{BA}, \label{eq:coordinate_rotations_l}
\\
m'-m_0 &=(l - l_0)\sin\mathrm{BA} + (m - m_0)\cos\mathrm{BA}.
\label{eq:coordinate_rotations_m}
\end{align}
The FWHM of the Gaussian along each principal axis is related to its corresponding standard deviation by
\begin{equation}
\mathrm{FWHM} = 2\sqrt{2\ln 2}\,\sigma \approx 2.355\,\sigma,
\end{equation}
which yields the beamwidth in the two principal directions by substituting $\sigma_{l}$ and $\sigma_{m}$, respectively. This model, while simple and not capable of capturing complex beam features such as sidelobes or asymmetries, provides a useful baseline for evaluating more advanced models.

\subsection{Airy disc beam model}

The Airy disc model (\citealt{Airy1835}) describes the diffraction pattern produced when EM waves, such as radio waves, pass uniformly through a circular aperture. 
This pattern is characterised by a central bright region (the Airy disc) surrounded by concentric rings of decreasing intensity. The size of the Airy disc, and hence the angular resolution of the antenna, is governed by the ratio of the observing wavelength to the aperture diameter. The ideal Airy pattern is a structured intensity distribution with a bright central peak and alternating dark and bright rings, resulting from constructive and destructive interference.

To better match the slightly asymmetric beam shapes seen in practice, we adopt a two-dimensional elliptical Airy disc model (\citealt{EllipticalAiry}) with rotational freedom. This modified version allows for ellipticity and rotation of the major axis, thereby providing flexibility in modelling beam asymmetries introduced by structural or instrumental effects.

The intensity pattern of the model is given by
\begin{align}
F_\mathrm{A}(l, m) = 
\left[ \frac{2\, J_1\bigl(\pi\, r_A(l, m) \bigr)}{\pi\, r_A(l, m)} \right]^2,
\label{eq:Airy}
\end{align}
where $J_1$ is the Bessel function of the first kind of order one and $r_A$ is a normalised radial coordinate that accounts for ellipticity and rotation.
The radial distance is defined as
\begin{align}
r_A(l, m) = \frac{\sqrt{(l' - l_0)^2 + q\, (m' - m_0)^2}}{r_{\text{major}} / \rho},
\end{align}
where the primed coordinates are obtained by rotating about the beam centre $(l_0, m_0)$ through Eqs. (\ref{eq:coordinate_rotations_l}) and (\ref{eq:coordinate_rotations_m}).

Here, $r_{\text{major}}$ and $r_{\text{minor}}$ are the semi-axes, $q = (r_{\text{major}}/r_{\text{minor}})^2$ 
encodes the ellipticity. The constant $\rho \approx 0.61$, derived from the first zero of the Bessel function, normalises the radial coordinate so that the first null occurs at $r_{\text{major}}$ along the major axis. This model combines the diffraction-limited structure of a circular aperture with ellipticity and rotation, capturing beam asymmetries better than symmetric Gaussian or Airy disc beam models alone.

\subsection{JimBeam model}

The JimBeam model, as described by \citet{deVilliers2022}, is built on the cosine aperture taper originally derived by \citet{CondonRansom2016}. A uniformly cosine-tapered aperture illumination yields a beam pattern of the form
\begin{equation}
    F_{\mathrm{J}}(l,m) = \left[\frac{\cos[\pi r_J(l,m)]}{1 - 4r_J(l,m)^2}\right]^2,
\end{equation}
The model is parameterised by frequency-dependent pointing centres $(l_0, m_0)$ and beam widths in both directions $(l_{\mathrm{FWHM}},\,m_{\mathrm{FWHM}})$, determined from holographic measurements. The elliptical beam radius, $r_J$, is defined by
\begin{equation}
    r_J^2(l,m) = r_0^2 \left[\left(\frac{l - l_0}{l_{\mathrm{FWHM}}}\right)^2 +
              \left(\frac{m - m_0}{m_{\mathrm{FWHM}}}\right)^2 \right],
\end{equation}
where $r_0 = 1.1889647809329453$.

Beam predictions are obtained using the publicly available \texttt{katbeam} library\footnote{\url{https://github.com/ska-sa/katbeam}}.
In this model, the Stokes~$I$ power beam is composed of the co-polarisation response only, with cross-polarisation contributions set to zero.

\subsection{Eidos beam model}

An alternative approach decomposes simulated or measured beam patterns into principal 
components (eigenbeams) rather than assuming an analytical form. The beam is represented 
as a sum of basis functions, with dominant modes capturing most of the structure. Instead 
of data-driven eigenmodes, one can use a fixed analytical basis such as Zernike polynomials, 
which are particularly suited to circular apertures: they are orthogonal, complete, defined 
on the unit disc, and encode physically meaningful aperture aberrations. This approach is 
adopted by the Eidos model.

The Eidos algorithm \citep{Iheanetu2019} represents the current state-of-the-art for modelling MeerKAT primary beams compactly. It fits Zernike polynomials \citep{Zernike2011} to holographic measurement data, capturing the dominant spatial modes of the beam with frequency-specific coefficients. Because the primary beam scales with frequency across the L band, the model must independently fit or reweight the Zernike decomposition at each frequency channel.

In the Eidos model\footnote{\url{https://github.com/ratt-ru/eidos}}, each frequency channel is decomposed using 100 Zernike polynomials, retaining only the 20 most dominant modes by power contribution. This provides a compact yet accurate representation, particularly for the Jones matrix diagonal elements. The model averages results over three antennas, which limits its ability to represent inter-antenna variations. Nevertheless, it is already suitable for Stokes I primary beam correction and direction-dependent calibration \citep{Asad2021}.

\begin{figure}
    \centering
    \includegraphics[width=\linewidth]{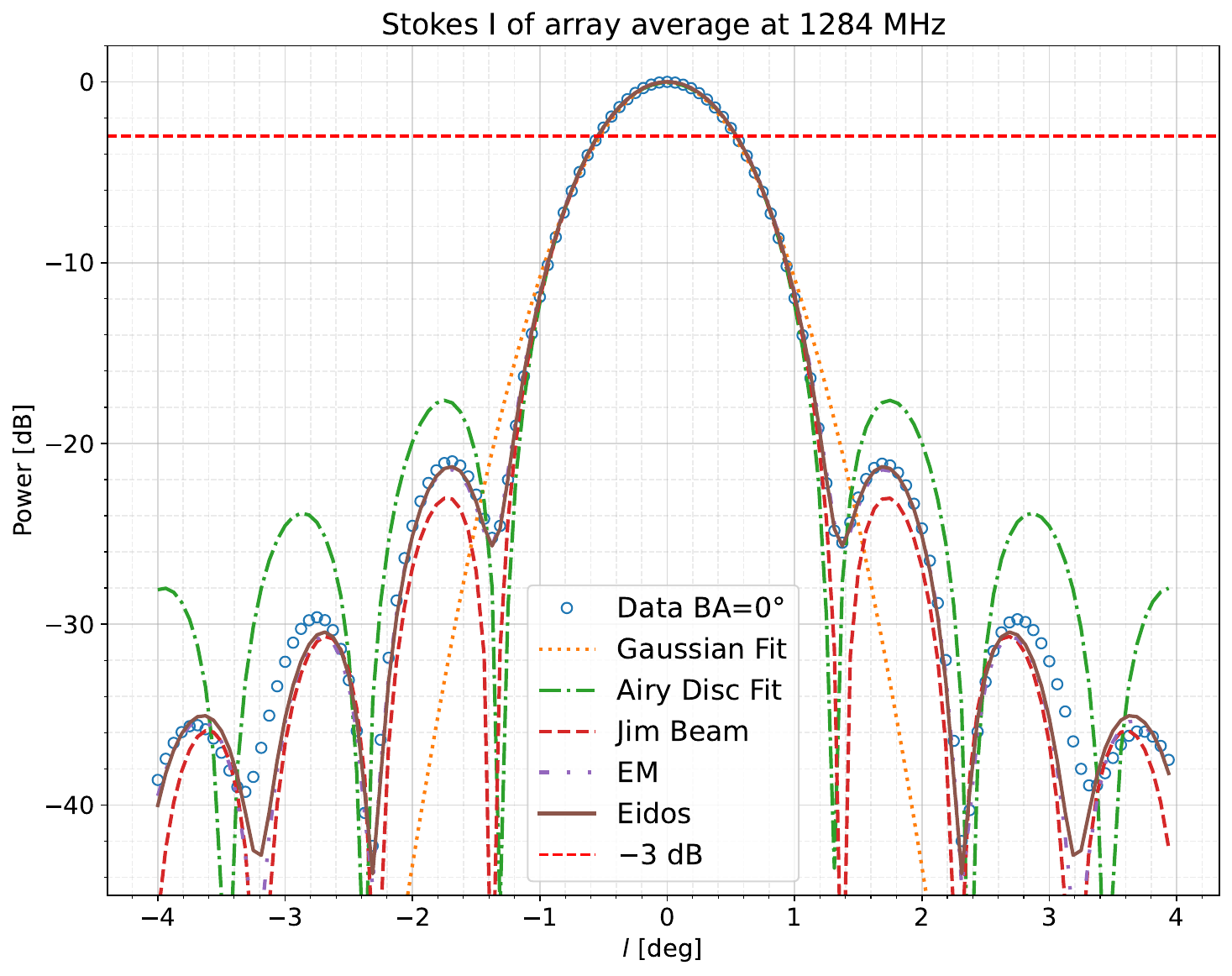}
    \caption{Comparison of beam pattern models for MeerKAT array average at 1284 MHz and angular offset at BA\,=\,$0\degr$. Holographic measurements (blue circles) are compared against Gaussian (orange dots), Airy disc (green dash-dots), Jim Beam (red dashes), EM (purple dash-dot-dots), and Eidos (brown line) models. The horizontal line marks the -3 dB FWHM level.}
    \label{fig:models_comparison_FWHM}
\end{figure}

Fig.~\ref{fig:models_comparison_FWHM} shows clear differences in how well each model captures the measured antenna response at 1284 MHz. The Gaussian fit approximates the main lobe but lacks sidelobes, while the Airy disc fit captures the oscillatory structure but overestimates sidelobe levels. Although it may perform better at some frequencies, the JimBeam  is shown to underestimate the level of the first sidelobe at the centre of the band, whereas the Eidos model is more capable of matching both the main lobe and first sidelobe. The JimBeam model does not parameterise the sidelobe level which changes with frequency. This demonstrates that while simpler analytical models capture general beam characteristics, higher-dimensional models like Eidos can provide more accurate representations of the full antenna response.

\section{Non-parametric beam modelling with NIFTy}
\label{sec:nifty-model}

We propose a different approach. Rather than describing the beam pattern at each frequency independently, we exploit the fact that the spatial structure of the beam is, to a very good approximation, common to the entire L band, with the frequency dependence concentrated in a small number of low-dimensional geometric transformations. Our approach (GAMBIT model) is based on a non-parametric Bayesian formulation grounded in IFT, a theoretical framework that extends Bayesian inference to continuous fields. Building on this foundation, we employ NIFTy, a computational library specifically designed for Bayesian inference in high-dimensional or function-space problems. NIFTy provides tools to estimate statistical properties of complex posterior distributions from noisy and incomplete data, and has been demonstrated to scale efficiently to problems involving half a billion parameters. At the core of NIFTy are advanced Gaussian process models and variational inference techniques. These algorithms are designed to handle non-linear, high-dimensional inference problems with both scalability and computational efficiency.

We made use of the JAX-based (Just After eXecution; \citealt{jax2018}) implementation of NIFTy (NIFTy.re; \citealt{niftyre}), which is designed for high-performance inference in high-dimensional spaces. By leveraging just-in-time compilation and automatic differentiation, this implementation provides significantly faster computations.

\subsection{MeerKAT beam modelling}

The Stokes I beam exhibits a cone-like structure (Fig.~\ref{fig:3D_StokesI}), as the main lobe narrows towards higher frequencies. However, this trend is more nuanced than simple self-similar scaling.

\citet{deVilliers2023} noted that MeerKAT's pointing performance suffers from pointing error because current calibration algorithms do not account for frequency-dependent beam variations. \citet{CottonMauch2021} 
demonstrated that even small pointing deviations introduce significant systematic errors 
in imaging and flux reconstruction.

Our analysis (see Appendix~\ref{App:ShiftParameterComparison}) further supports these findings and suggests that a simple frequency scaling is insufficient to fully capture the spectral structure of the beam. Instead, we observe systematic frequency-dependent shifts in the beam centroid, indicative of residual pointing-related effects that vary across the band. As illustrated in the upper panel of Fig.~\ref{fig:CenterOfMass}, the centroid offsets, $\Delta l_0$ and $\Delta m_0$, exhibit only weak mutual correlation and can therefore be modelled independently. This behaviour is consistent with the interpretation that frequency-dependent pointing errors contribute non-trivially to the apparent beam motion and cannot be adequately represented by a purely self-similar scaling model.

To capture both effects within a unified framework, we extend the model to include three latent parameters per frequency channel: a scale factor, $s_f$, and two spatial shift parameters, $\Delta l_{0,f}$ and $\Delta m_{0,f}$, accounting for offsets from the centre. The scale factor rescales the spatial beam field at each frequency, encoding the contraction or expansion of the beam with wavelength, whereas the shift parameters account for frequency-dependent translations of the beam centre. This formulation enables the simultaneous inference of a generalised, frequency-independent beam shape together with frequency-dependent geometric transformations, resulting in a compact, generative, and physically interpretable description of the beam’s spectral variations.

\subsection{Hierarchical Bayesian framework}

We developed a hierarchical Bayesian model to reconstruct the primary beam of MeerKAT across L band covering 856 to 1712\,MHz. The model decomposes the frequency-dependent beam into a base pattern shared across all frequencies and frequency-specific geometric transformations.

\subsubsection{Model structure}

The observed holographic data at frequency $f$ and angular position $(l, m)$ is denoted as $d_f(l, m)$, where $l$ and $m$ are angular offsets from the beam centre. We model these observations as
\begin{equation}
     d_f(l, m) = M_f(l,m) + n_f(l, m),
\end{equation}
where $M_f(l, m)$ is the model prediction and $n_f(l, m)$ represents measurement noise.

The frequency-dependent beam pattern is constructed by applying geometric transformations to a base beam
\begin{equation}
    M_f(l,m) = T(B(l,m); s_f, \Delta l_{0,f}, \Delta m_{0,f}),
\end{equation}
where $B(l,m)$ --- the base beam --- is represented as a correlated Gaussian field ($\mathrm{CGF}$) using the NIFTy framework shared across all frequencies, $s_f$ is the scale factor at frequency $f$, and $\Delta l_{0,f}$ and $\Delta m_{0,f}$ are the spatial shifts of $(l_0, m_0)$ at frequency $f$.

The transformation operator, $T$, applies these geometric operations sequentially
\begin{equation}
    T(B(l,m); s_f, \Delta l_{0,f}, \Delta m_{0,f}) = B(s_f (l - \Delta l_{0,f}), s_f (m - \Delta m_{0,f})).
\end{equation}
This formulation ensures the beam is first shifted to correct the centre position, then scaled about the corrected centre, providing a physically intuitive decomposition of the frequency-dependent beam variations. The scale transformation is implemented via coordinate remapping with bilinear interpolation using \texttt{map\_coordinates} from
\texttt{jax.scipy.ndimage}~\citep{jax2018}, which provides automatic differentiation support.

The base beam, $B(l,m)$, itself is modelled as a correlated Gaussian field (CGF) defined on a 2D spatial grid. In NIFTy, under the assumption of  homogeneity, such fields are characterised by their power spectrum, $P(k)$ \footnote{\href{https://ift.pages.mpcdf.de/nifty/user/notebooks_re/4_correlated_field_model.html}{NIFTy CGF documentation}}, which specifies the amplitude of correlations at each spatial frequency, $k$ (for more details see \citealt{arras2019nifty} and \citealt{Arras2022}), with physically motivated priors placed on its hyperparameters. This approach provides a regularised yet highly flexible representation: smooth at large scales by construction, but capable of encoding fine spatial detail where supported by the data. The power spectrum is parametrised by a small set of hyperparameters which will be described in the next section.

\subsubsection{Prior specification}

All prior hyperparameters were calibrated in two steps: first, initialisation from a model-free analysis of the data ($\lambda/D$ scaling for zoom factors, centre-of-mass analysis for shifts; see Appendix~\ref{App:ShiftParameterComparison}); and second, validation via prior predictive checks, confirming that the resulting prior ensemble spans a broad range of physically plausible beam features. To reduce cost, this was first performed on 100 equi-spaced frequency channels before being applied to the full inference. Calibrating the CGF hyperparameters this way is important, since an overly narrow prior can suppress the small-scale beam structure the model is designed to capture.

Each hyperparameter of the CGF, listed in Table~\ref{tab:cgf_hyperparameters}, controls a distinct aspect of the field's amplitude and correlation structure. With the exception of \texttt{offset\_mean} (fixed to zero) and \texttt{loglogavgslope} (assigned a Gaussian prior), each hyperparameter is given as a (mean, standard deviation) pair that parametrises a log-normal prior. The resulting priors give the base beam model enough freedom to accommodate genuine localised or non-power-law structure.

\begin{table*}[h]
\centering
\caption{Prior distributions of the model hyperparameters. The upper block gives the hyperparameters of the CGF prior on the base beam $B(l, m)$; the lower block gives the priors on the frequency-dependent scale factors and spatial shifts.}
\label{tab:cgf_hyperparameters}
\renewcommand{\arraystretch}{1.15}
\begin{tabular}{@{}l l p{11cm}@{}}
\hline\hline
Hyperparameter & Prior distribution & Description \\
\hline
\texttt{offset\_mean}     & $0.0$ (fixed)                   & Zero-mode level of $B(l, m)$. \\
\texttt{offset\_std}      & $\mathrm{LogNormal}(0.2, 0.9)$  & Prior width around the offset. \\
\texttt{fluctuations}     & $\mathrm{LogNormal}(0.3, 1.0)$  & Overall amplitude of the field's variations about its offset. \\
\texttt{loglogavgslope}   & $\mathcal{N}(-2.5, 0.2)$        & Average slope of $\log P(k)$ vs.\ $\log k$, setting the field's smoothness. \\
\texttt{flexibility}      & $\mathrm{LogNormal}(0.4, 0.7)$  & How strongly the power spectrum varies beyond the power-law trend. \\
\texttt{asperity}         & $\mathrm{LogNormal}(0.1, 3.0)$  & How rough those deviations are allowed to be from one $k$-bin to the next.\\

\multicolumn{3}{l}{} \\
\hline
$s_f$                     & $\mathrm{LogNormal}(1.0, 0.34)$ & Isotropic scale factor at frequency $f$ \\
$\Delta l_{0,f}$          & $\mathcal{N}(0.0, 0.1)$         & Horizontal spatial shift at frequency $f$ \\
$\Delta m_{0,f}$          & $\mathcal{N}(0.0, 1.2)$         & Vertical spatial shift at frequency $f$ \\
\hline
\end{tabular}
\end{table*}

We adopt log-normal and Gaussian priors for the scale factors $\{s_f\}$ and shift parameters $\{\Delta l_{0,f}, \Delta m_{0,f}\}$, respectively, with the hyperparameter values listed in Table~\ref{tab:cgf_hyperparameters}. The log-normal prior on $s_f$ ensures positivity, while allowing for smooth variation around unity, and the asymmetry $\sigma_m \gg \sigma_l$ reflects the empirical centre-of-mass behaviour discussed in Appendix~\ref{App:ShiftParameterComparison}, where the vertical shift is systematically larger than the horizontal one.

\subsubsection{Noise model}

We assume independent Gaussian noise at each frequency with variance $\sigma^2$
\begin{equation}
P(d_f | M_f, \sigma) = \mathcal{N}(n_f, \sigma^2 \mathrm{I}).
\end{equation}
The noise level was estimated using the median absolute deviation (MAD; \citealt{MAD_estimator}) applied to edge regions of the holographic measurements, where beam power is negligible. For each frequency channel, we extracted pixels from the outer 10\% of the image borders. The MAD estimator, given by $\sigma_{\text{MAD}} = 1.4826 \times \text{median}(|x_i - \text{median}(x)|)$, where $x = \{x_1, x_2, \ldots, x_N\}$ represents the set of pixel values in the border region, is preferred over standard deviation for noise estimation in the presence of signal contamination.

This yields a mean noise estimate of $\sigma_{\mathrm{MAD}} \approx 1.57 \times 10^{-4}$. 
Fig.~\ref{fig:Noise analysis} compares estimates from three regions: image edges, outer ring, and corners. Corner regions, farthest from the beam centre, provide the most reliable estimates due to minimal beam contamination. Despite lying closer to the sidelobes, edge and outer-ring estimates remain consistent with corner values, indicating 
negligible signal leakage. Given this agreement and the modest frequency dependence observed, we adopt a single noise level for all frequencies.

However, this pure measurement noise does not account for systematic uncertainties inherent in holographic observations. To provide a more realistic uncertainty budget, we adopted a conservative noise level of $\sigma = 8 \times 10^{-4} \approx 5\sigma_{\text{MAD}}$. This inflation factor accommodates: (i) residual radio frequency interference (RFI) and calibration errors, (ii) model inadequacy in capturing fine beam structure beyond our model, and (iii) environmental variations (e.g. atmospheric and ground spillover) during observations.

A sensitivity sweep over $\sigma \in [2 \times 10^{-4},\, 2 \times 10^{-3}]$ on the inference of 30 representative frequency channels shows that both the residual variance and the residual spatial correlation are invariant to the assumed noise, indicating that they are set by a model-limited structural floor rather than by the likelihood weighting.

\subsubsection{Posterior distribution and inference}

The posterior distribution over all parameters is
\begin{equation}
    P(\xi | \{d_f\}) \propto P(\{d_f\} | \xi) \: P(\xi),
\end{equation}
where the parameter vector is
\begin{equation}    
\xi = \{\mathrm{CGF}_{\text{params}}, \{s_f\}, \{\Delta l_{0,f}\}, \{\Delta m_{0,f}\}\}.
\label{eq:xi}
\end{equation}
The likelihood factorises over frequencies
\begin{equation}
    P(\{d_f\} | \xi) = \prod_{f=1}^{N_{\text{freq}}} P(d_f | M_f(\xi), \sigma_f).
\end{equation}

Posterior inference in NIFTy.re is performed using variational inference (\citealt{knollmuller2019mgvi,GeometricVariationalInference}), which provides an efficient approximation to the posterior distribution in high-dimensional field-inference problems. Rather than sampling the full posterior directly, variational inference approximates it with a simpler, parametrised distribution $q(\xi) = \mathcal{N}(\xi - \mu, \mathbf{C})$---here a Gaussian with mean $\mu$ and covariance $\mathbf{C}$---and determines the parameters of $q$ by minimising the Kullback--Leibler (KL) divergence between the approximation and the true posterior $P(\xi|\{d_f\})$
\begin{equation}
    \mathrm{KL}(q(\xi) \| P(\xi|\{d_f\})) = \mathbb{E}_q[\log q(\xi)] - \mathbb{E}_q[\log P(\xi|\{d_f\})].
\end{equation}
The variational inference procedure iteratively refines both the base beam structure and the frequency-dependent transformation parameters until convergence is achieved. Posterior samples are then drawn from the approximate distribution $q(\xi)$ to quantify uncertainties on all inferred quantities.

To suppress the periodic boundary condition implicit in the correlated field model, we enlarge the domain of the base-beam correlated field by 20\% on each side relative to the data grid, keeping the pixel scale fixed. Only the central region matching the data extent is passed to the likelihood, and the outer boundary is discarded after inference. This prevents edge artefacts near the data boundaries, ensuring that spatial correlations are properly
constrained without artificial discontinuities.

\subsection{Results}

\begin{figure*}
    \centering
    \includegraphics[width=\textwidth]{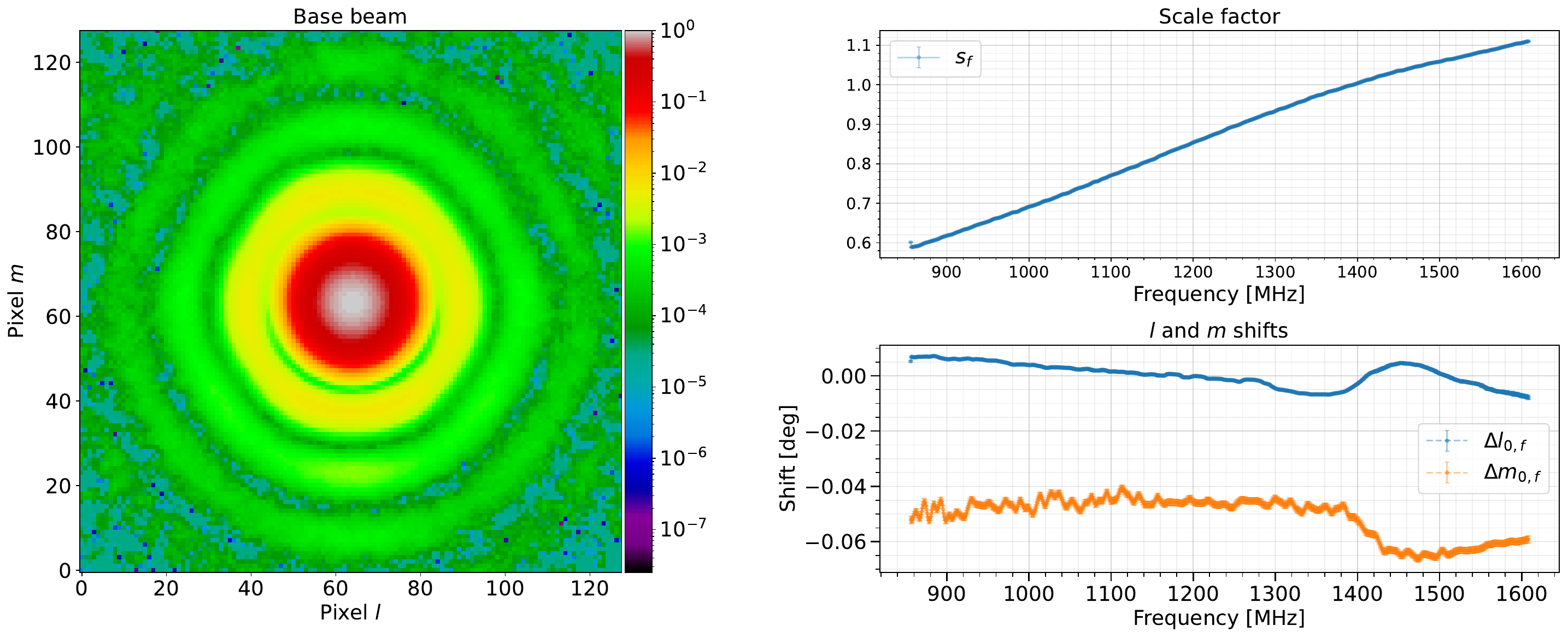}
    \caption{Posterior mean estimates from the hierarchical Bayesian beam reconstruction model for MeerKAT array average. \textit{Left:} Base beam pattern $B(l,m)$ shared across all L band frequencies. \textit{Top right:} scale factor, $s_f$, as a function of frequency, exhibiting monotonic increase from $\sim$0.6 at 900~MHz to $\sim$1.1 at 1700~MHz, capturing the frequency-dependent beam size variation. \textit{Bottom right:} Spatial shift parameters $\Delta l_{0,f}$ (blue) and $\Delta m_{0,f}$ (orange) showing systematic frequency-dependent beam centre translations. Error bars represent posterior standard deviations. The smooth functional forms validate the model's ability to capture coherent frequency-dependent beam variations.}
    \label{fig:beam_scaling_shift_mean}
\end{figure*}

\begin{figure}
    \centering
    \includegraphics[width=0.95\linewidth]{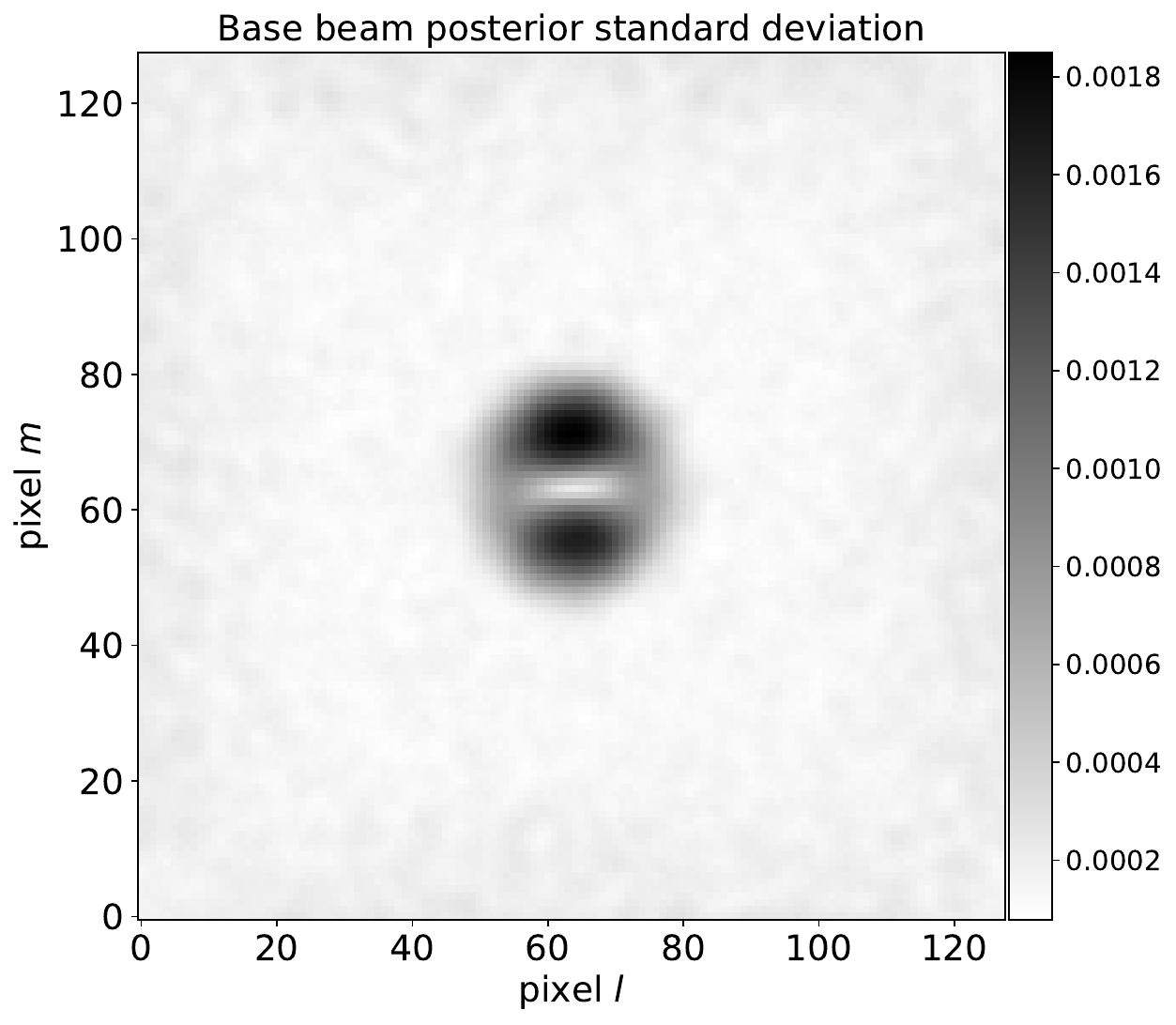}
    \caption{Posterior standard deviation of the inferred base beam showing pixel-wise reconstruction uncertainty.}
    \label{fig:Base_beam_std}
\end{figure}

Using the L band holographic Stokes I data from the MeerKAT array average as input to GAMBIT model, we obtained posterior estimates of the beam pattern and its frequency-dependent geometric transformations. Fig.~\ref{fig:beam_scaling_shift_mean} presents the posterior mean of the inferred base beam structure along with the corresponding means of the scale factor and spatial shift parameters across frequency. The posterior uncertainty of the base beam shown in Fig.~\ref{fig:Base_beam_std} is strongly non-uniform: it peaks near the first sidelobe and drops by roughly an order of magnitude in the outer regions.

The right top panel of Fig.~\ref{fig:beam_scaling_shift_mean} shows the inferred scale factor as a function of frequency across the L band. The scale factor increases monotonically from approximately 0.6 at the lower edge to 1.1 at the upper edge, confirming systematic beam contraction at higher frequencies consistent with diffraction physics. The right bottom panel reveals the frequency dependence of the spatial shift parameters. The $\Delta l_{0,f}$ parameter (blue) shows a systematic trend, varying from $+0.01$\,deg at low frequencies, crossing zero near 1200~MHz, reaching a minimum of approximately $-0.01$\,deg near 1450~MHz, and returning towards zero at the band edge. The $\Delta m_{0,f}$ parameter (orange) exhibits complementary behaviour, varying between approximately $-0.04$ and $-0.06$\,deg with a broad maximum around 1150~MHz. These smooth, well-constrained frequency dependencies (indicated by the small error bars) suggest systematic effects that cause the beam centre to shift across the band.

The combination of scale and shift parameters allows the model to separate intrinsic beam structure from geometric transformations, yielding a compact and physically interpretable description of the measured data. Overall, the posterior mean beam and frequency-dependent transformation parameters represent the most probable underlying beam behaviour given the holographic measurements and the Bayesian prior assumptions. This provides a robust foundation for downstream calibration and imaging tasks where accurate knowledge of the primary beam is essential.

\begin{figure}
    \centering
    \includegraphics[width=\linewidth]{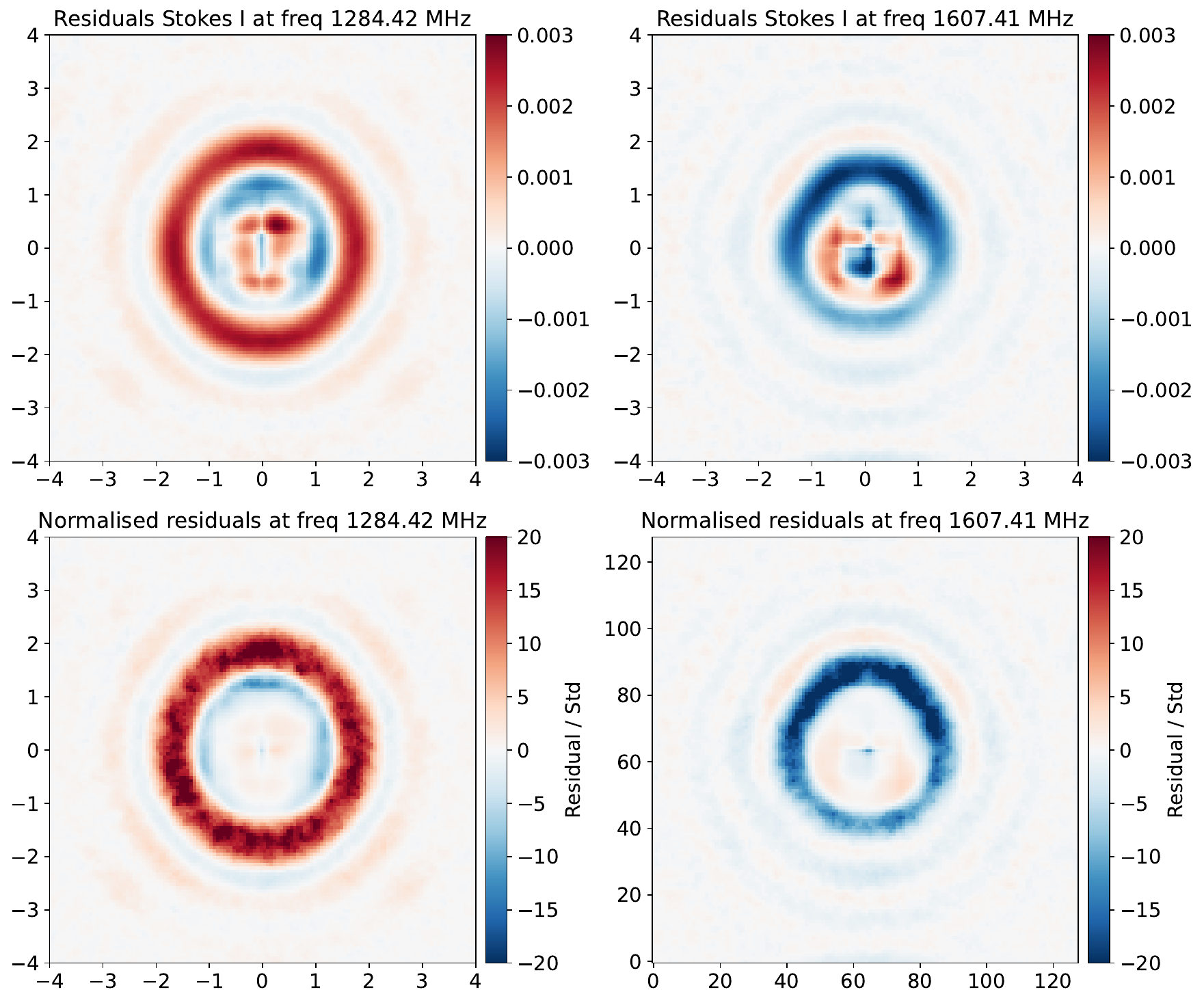}
    \caption{Example residual maps (data minus model) at two representative frequencies (1284.42~MHz, left; 1607.41~MHz, right), demonstrating the reconstruction accuracy of the model. \textit{Top row:} Residuals in normalised beam units, with typical amplitudes of $\pm 0.003$. \textit{Bottom row:} Residuals normalised by the posterior standard deviation of the model at each pixel.}
    \label{fig:residual_beam_scaling_shift_mean}
\end{figure}

Fig.~\ref{fig:residual_beam_scaling_shift_mean} shows two example residual maps (data minus model) at 1284.42~MHz and 1607.41~MHz. The raw residuals (top row) range from $-0.003$ to $+0.003$—sub-percent deviations from the measured beam. The bottom row shows the same residuals normalised by the posterior standard deviation of the model. Although the raw residuals are small ($<0.3\%$ of the peak), the normalised maps reveal systematic frequency-dependent structure at the level of several standard deviations near the first sidelobe. The systematic ring-like and asymmetric patterns visible in both representations arise from small structural differences relative to the single-frequency base beam (see Appendix \ref{app:ZernikeBaseBeam}) that cannot be fully captured by the frequency-dependent scale and shift transformations alone, motivating future extensions.  A quantitative comparison with other beam models is presented in the following section.

\subsection{Comparison to alternative models}

We evaluated model accuracy using mean squared deviation: $\text{MSD}(f) = \langle (M_f - d_f)^2 \rangle_{\text{pixels}}$, 
where $M_f$ and $d_f$ are model and data at frequency $f$. This quantifies the mean squared residual (e.g. Fig.~\ref{fig:residual_beam_scaling_shift_mean}) averaged over the spatial domain. We also define the band-integrated metric $\Sigma = \sum_f \text{MSD}(f)$.

\begin{figure*}
    \centering
    \includegraphics[width=0.85\textwidth]{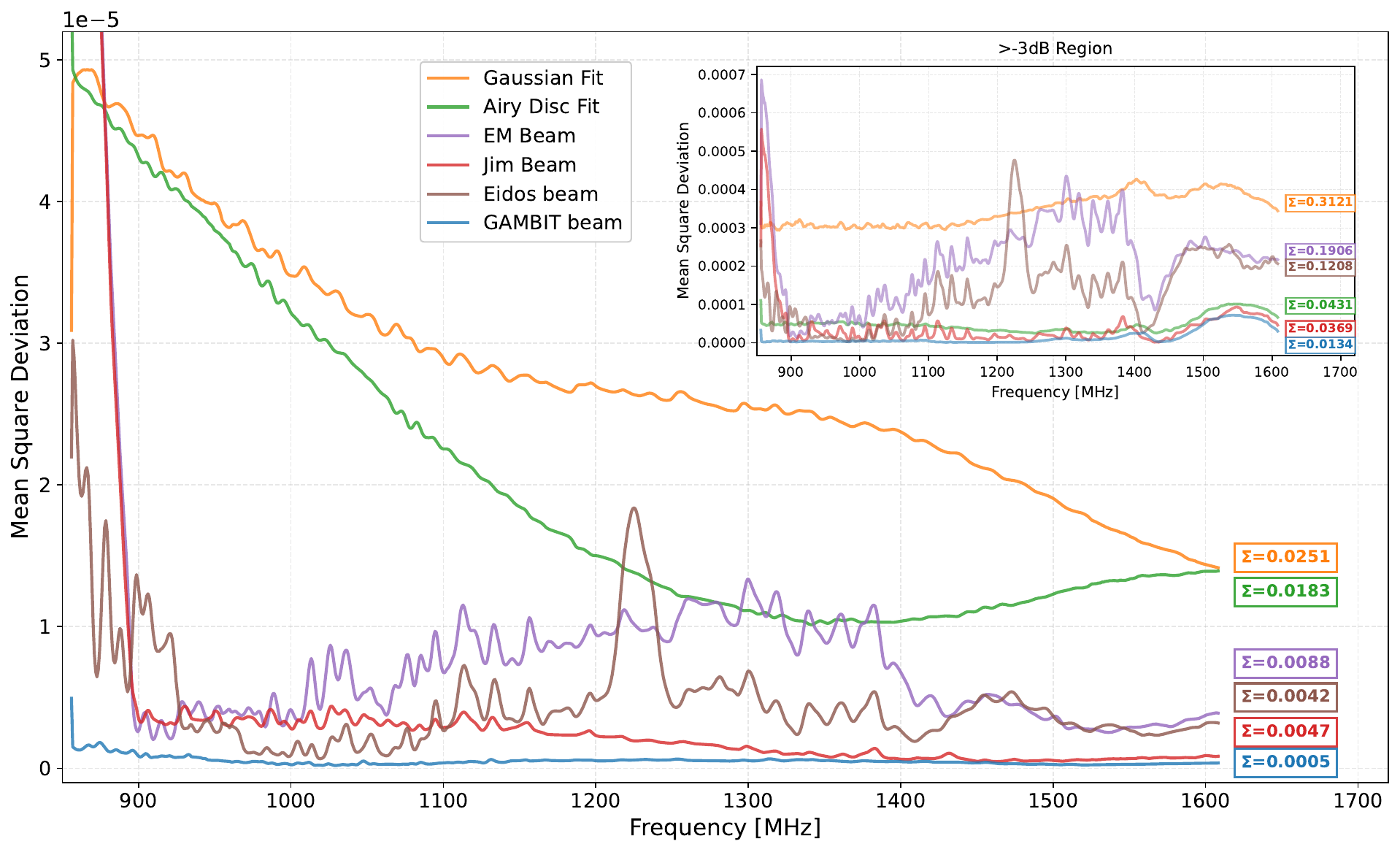}
    \caption{Mean squared deviation comparison across beam models versus frequency. \textit{Main plot:} Full-field MSD. Our GAMBIT model (blue) achieves the lowest deviation across the L band, outperforming alternatives by over an order of magnitude. Eidos (brown) is next best, followed by JimBeam (red), EM (purple), Airy disc (green), and Gaussian (orange). \textit{Inset:} MSD for the $>-3$\,dB main lobe region only. Our model (blue) remains the most accurate. Notably, the analytical JimBeam outperforms Airy disc, Eidos, and EM in this region, while Gaussian shows the largest main-lobe errors.} 
    \label{fig:MSD_different_models}
\end{figure*}

We compute the MSD between each beam model and the holographic measurements across the L band (856--1607\,MHz); results are shown in
Fig.~\ref{fig:MSD_different_models}. In the full-field comparison (main panel), GAMBIT (blue, $\Sigma = 0.0005$) outperforms all alternatives by nearly an order of magnitude. Eidos (brown, $\Sigma = 0.0042$) and JimBeam (red, $\Sigma = 0.0047$) are the closest competitors, with Eidos marginally better in the band-averaged sense but showing elevated deviations around 1100--1300\,MHz. The EM simulation (purple, $\Sigma = 0.0088$) performs worse than both, reflecting known chromatic limitations of idealised EM models \citep{Iheanetu2019,Asad2021,deVilliers2022}. The Airy disc (green, $\Sigma = 0.0183$) and Gaussian (orange, $\Sigma = 0.0251$) show progressively larger deviations confirming the inapplicability of these functional forms. We note that
several models exhibit elevated deviations at the lowest frequencies ($< 900$\,MHz), the origin of which has not been investigated at the time of writing.

The inset shows the MSD restricted to the main-lobe region, where the ranking changes notably. GAMBIT remains the best model ($\Sigma = 0.0134$), but JimBeam ($\Sigma = 0.0369$) now outperforms both the Airy disc ($\Sigma = 0.0431$) and Eidos ($\Sigma = 0.1208$), the latter showing substantially higher main-lobe deviations. The JimBeam (MKAT-AA-L-JIM-2020) model is used directly from the  \texttt{katbeam} package, where pointing offsets and beamwidths are based on averaged samples over 50\,MHz intervals. We found (not shown) that re-fitting the JimBeam on a per-channel basis yields slightly lower MSD values, as the finer frequency sampling better resolves the spectral variation in the beam size and centroid position.

The upper end of the L band above 1607 MHz, is excluded from this comparison to avoid regions where the beam exhibits anomalous behaviour, including waveguide mode excitation at the high-frequency end. Within the remaining frequency range, our results demonstrate that explicitly modelling frequency-dependent geometric transformations through a hierarchical, data-driven Bayesian framework yields a more accurate and robust representation of the MeerKAT primary beam than existing analytical, data-driven parametric or EM-based models.

\begin{figure}
    \centering
    \includegraphics[width=\linewidth]{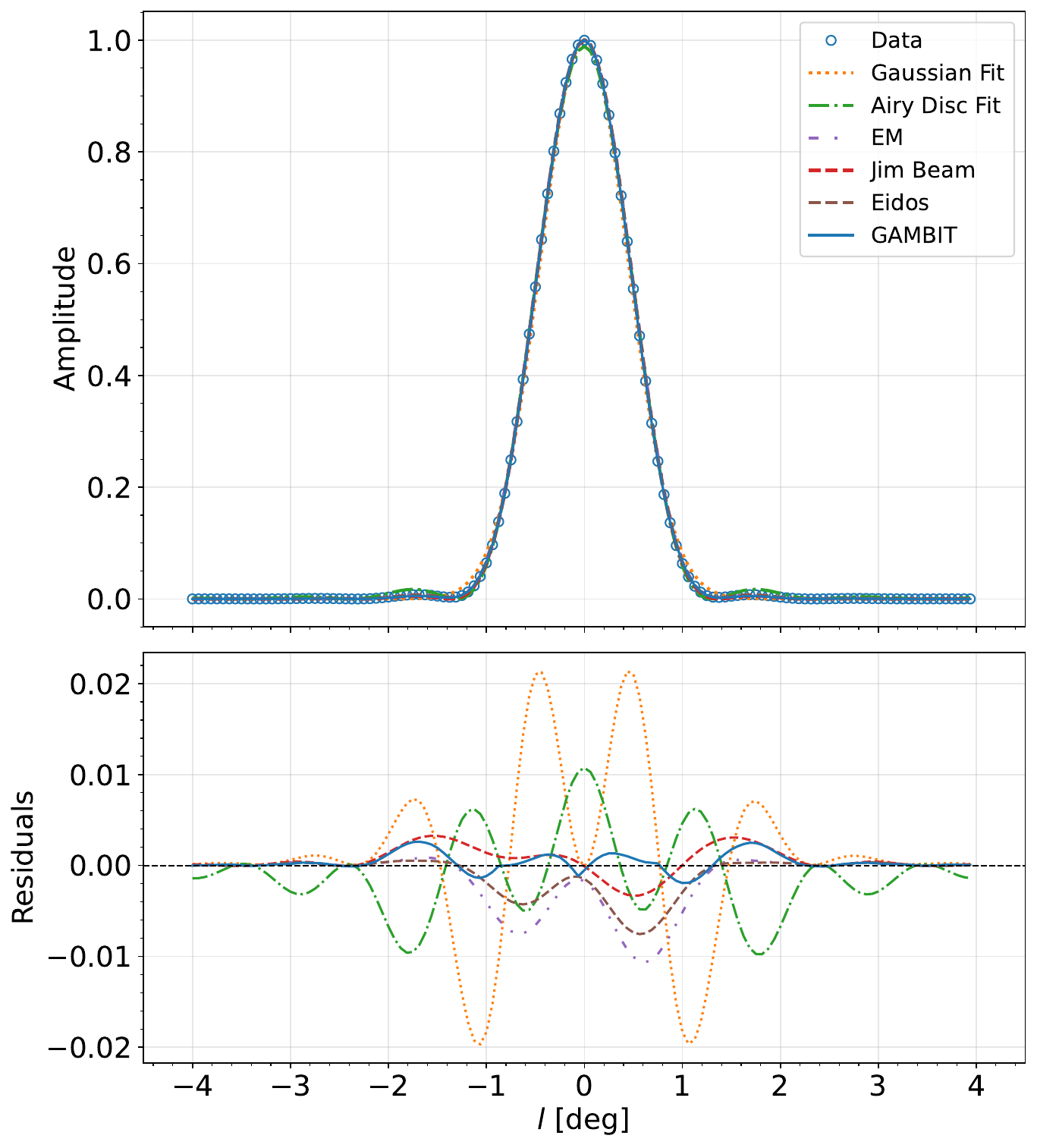}
    \caption{Beam pattern comparison for array average at 1282 MHz and BA\,=\,$0\degr$. Top: Data (blue circles) compared to Eidos (brown dashes) and our model reconstruction (solid blue line) in linear scale. Bottom: Residuals (data minus model prediction).}
    \label{fig:BeamComparisonWithResiduals}
\end{figure}

To see the main lobe differences in more detail, we directly compare one-dimensional cuts of the beam predicted by the different models with the measured beam at BA$\,=\,0\degr$. Fig.~\ref{fig:BeamComparisonWithResiduals} shows the beam profiles on a linear scale, together with the corresponding residuals in the lower panel. Our GAMBIT model reproduces the main lobe with high accuracy, resulting in small, well-behaved residuals near the beam centre. 

\section{Discussion of MeerKAT beam properties}
\label{sec:meerkat-beam}
We now turn to a discussion of the physical properties of the MeerKAT array-averaged beam, as obtained from our model and in comparison to the holographic beam data.

\subsection{Beam size and beam shape}

The inferred scale factor, $s_f$, encodes the frequency-dependent contraction of the beam. It exhibits a clear break at $f_{\rm bp} = 1391.4 \pm 2.5$\,MHz, well described by a continuous piecewise-linear fit on both sides. The slope decreases by a factor of $1.59$ across the break, from $7.92\times10^{-4}$\,MHz$^{-1}$ below $f_{\rm bp}$ to $4.98\times10^{-4}$\,MHz$^{-1}$ above it. A corresponding change in trend is also visible in the centroid shifts ($\Delta l_{0,f}$ and $\Delta m_{0,f}$) at the same frequency. Interpreting $f_{\rm bp}$ as the cut-off frequency of the feed's first higher-order circular-waveguide mode, TE$_{21}$, implies a waveguide throat diameter of $\approx 20.9$\,cm, in good agreement with \citet{deVilliers2021OMT}. This single measured quantity, with no additional free parameters, predicts the cut-off frequency of the fundamental TE$_{11}$ mode to be 838.8\,MHz, only 17.2\,MHz below the L band lower edge of
856\,MHz. This agreement -- obtained purely from the frequency dependence of a Bayesian beam-model latent parameters, without reference to the feed's mechanical drawings -- is, to our knowledge, the first demonstration that a generative beam model can recover a physical wave\-guide dimension directly from holographic data.

\begin{figure}
    \centering
    \includegraphics[width=\linewidth]{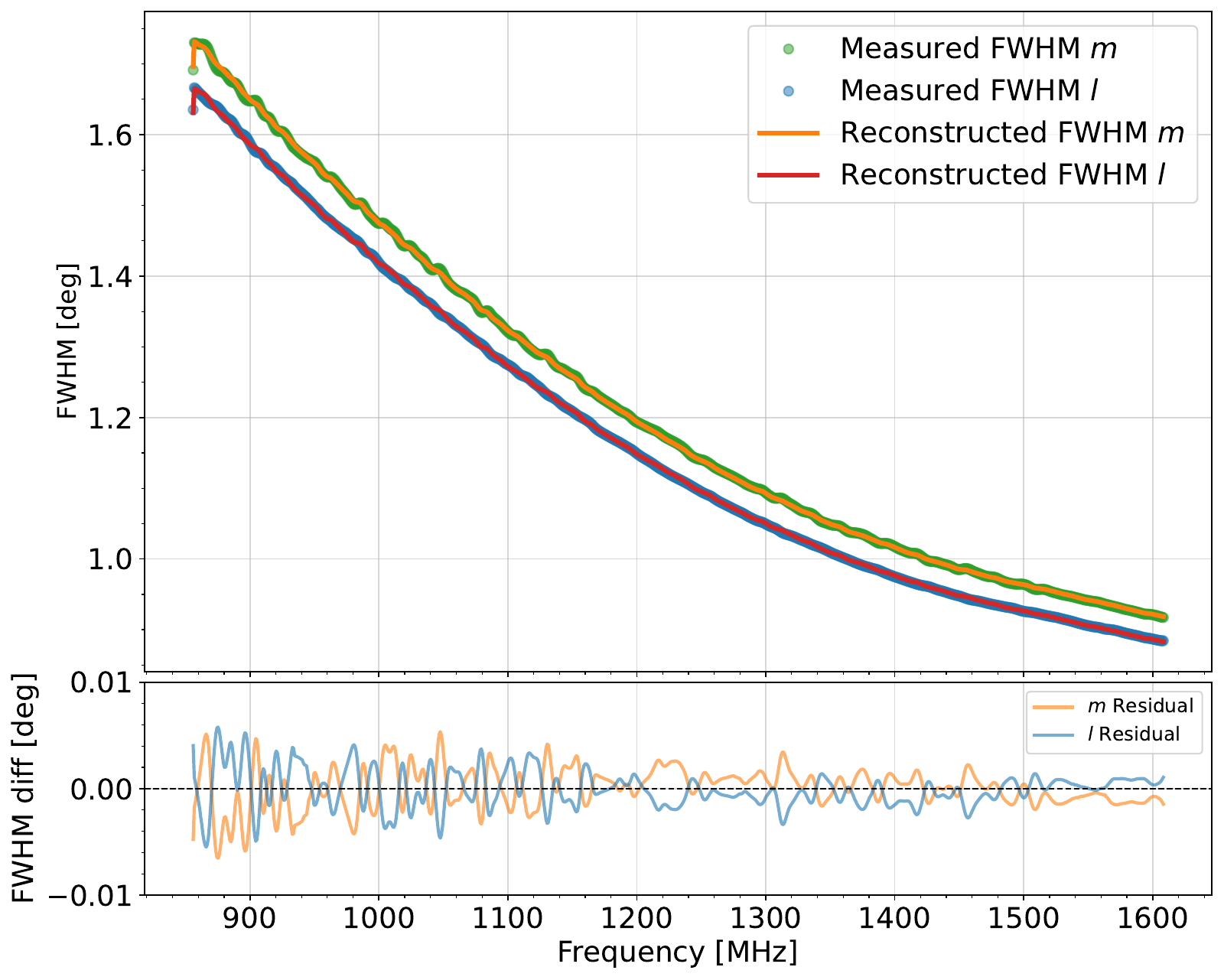}
    \caption{FWHM comparison between our model and holographic measurements. \textit{Top:} FWHM in $l$ (blue/red) and $m$ (green/orange) directions decreases from $\sim$1.7\,deg to $\sim$0.9\,deg across the L band. Measured values (light points) show oscillatory fine structure, while our model (solid lines) provides a smooth scaling. The $l$--$m$ offset indicates beam ellipticity. \textit{Bottom:} Residuals exhibit periodic oscillations of $\sim$1--2\% amplitude, representing fine-scale frequency variations beyond the smooth scaling.}
    \label{fig:FWHM_complete_model}
\end{figure}

The measured beam width exhibits small fluctuations superimposed on the overall FWHM trend (Fig. \ref{fig:FWHM_over_LBand}). To quantify these, we compare the FWHM of the reconstructed beams with that of the holographic measurements. Fig.~\ref{fig:FWHM_complete_model} compares FWHM values from holographic measurements with those from the GAMBIT model. The reconstructed FWHM values (red and orange curves) follow a smooth and monotonic decrease with frequency, consistent with diffraction-limited behaviour. In contrast, the measured FWHM values (blue and green dots) exhibit fine-scale oscillations of 0.01--0.02 degrees superimposed on the overall trend, representing 1--2\% variations at a given frequency. 

The smoothness of the GAMBIT reconstruction is a deliberate feature of the model that demonstrates successful separation of the beam into orthogonal components: the scale factor captures the dominant diffraction-limited scaling that pinpoints waveguide mode cut-off frequencies, while the shift parameters and the correlated field encode pointing offsets and spatial structure, respectively. 

Fourier analysis of the residuals reveals the physical origin of these oscillations. The FFT spectrum shows dominant peaks at 23.2, 13.5, and 50\,MHz, corresponding to cavity lengths of 3.5\,m, 2.0\,m, and 7.6\,m, respectively (using $L = c\tau/2$ for round-trip delay). These dimensions match key structures in the MeerKAT antenna: the feed-to-subreflector path ($\sim$3--4\,m), sub-to-main reflector ($\sim$10--20\,m). The oscillations thus may arise from standing waves generated by reflections at these structural boundaries, creating frequency-dependent interference patterns that modulate the effective beamwidth beyond the primary diffraction-limited scaling.

Unlike many analytical beam models, the holographic beam measurements and the GAMBIT reconstruction reveal a hexagonal-like structure to the sidelobe rings (Fig. \ref{fig:beam_scaling_shift_mean}) similar to optical telescopes with hexagonal apertures (\citealt{Sabatke2005}). This pattern arises by the Fourier transform from the geometry of the MeerKAT main reflector outline, due to the illumination taper extending at low power beyond the 13.5\,m nominal circular diameter.

\subsection{Zernike representation of the MeerKAT base beam}

The base beam, $B(l,m)$, is shared across all frequencies in our hierarchical model, with frequency dependence captured by the transformation parameters. To represent this base beam compactly, we employ a Zernike polynomial decomposition. This orthogonal basis provides two key advantages: the coefficients correspond to physically interpretable optical aberrations, and the continuous analytic form enables high-resolution beam reconstruction at an arbitrary spatial sampling.

The Zernike polynomials, $Z_n^m(\rho, \phi)$ \citep{BornWolf1999}, form a complete and orthogonal set of functions on the unit disc, characterised by radial order $n$ and azimuthal frequency $m$. The radial component is defined as
\begin{equation}
R_n^m(\rho) = \sum_{k=0}^{(n-m)/2} \frac{(-1)^k \, (n-k)!}{k! \, \left(\frac{n+m}{2}-k\right)! \, \left(\frac{n-m}{2}-k\right)!} \, \rho^{n-2k},
\end{equation}
where $n \geq 0$, $|m| \leq n$, and $n - |m|$ is even.

There are even and odd Zernike polynomials. The even Zernike polynomials are defined as
\begin{equation}
Z_n^{+m}(\rho, \phi) = R_n^m(\rho) \, \cos(m\phi) \quad \text{for } m \geq 0,
\end{equation}
and the odd Zernike polynomials as
\begin{equation}
Z_n^{-m}(\rho, \phi) = R_n^{-m}(\rho) \, \sin(-m\phi) \quad \text{for } m < 0.
\end{equation}
For $m = 0$, the polynomial is purely radial $Z_n^0(\rho, \phi) = R_n^0(\rho)$.

To apply the Zernike decomposition to the square beam map of dimensions $N_{l} \times N_{m}$ pixels, we first transformed the Cartesian pixel coordinates $(i, j)$ to the unit disc. For a pixel at position $(i, j)$ where $i \in [0, N_{l}-1]$ (row index) and $j \in [0, N_{m}-1]$ (column index), the normalised coordinates are
\begin{equation}
l_{ij} = \frac{j - N_{l}/2}{N_{l}/2}, \quad m_{ij} = \frac{i - N_{l}/2}{N_{l}/2},
\end{equation}
which are then converted to polar coordinates
\begin{equation}
\rho_{ij} = \sqrt{l_{ij}^2 + m_{ij}^2}, \quad \phi_{ij} = \arctan2(l_{ij}, m_{ij}).
\end{equation}
Pixels with $\rho_{ij} > 1$ lie outside the unit disc and are excluded from the analysis via a circular mask. This mapping ensures that the beam's central region (within the main lobe and inner sidelobes) is contained within the unit disc, where the Zernike polynomials are well defined and orthogonal. In the following, we use the Noll index $j = j(n,m)$ \citep{Noll76} to enable a matrix representation of the base beam decomposition and simple linear algebraic manipulation thereof. 

\begin{figure}
    \centering
    \includegraphics[width=\linewidth]{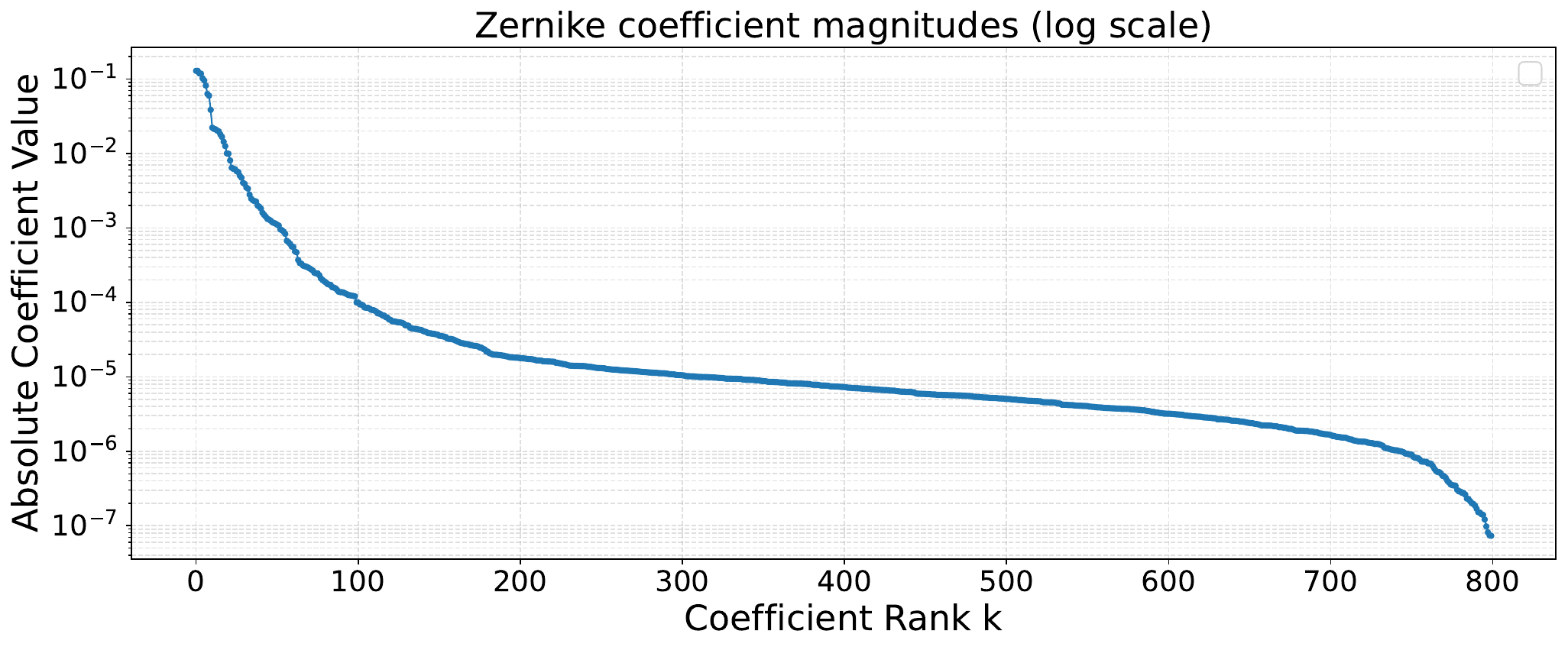}
    \caption{Zernike polynomial coefficients for the base mean beam of MeerKAT array average. The first 800 coefficient magnitudes are shown in logarithmic scale, with dominant contributions from low-order modes representing the primary beam aberrations.}
    \label{fig:zernike_modes_visualisation}
\end{figure}

Fig.~\ref{fig:zernike_modes_visualisation} shows the dominant Zernike coefficients for the inferred base beam $B(l,m)$. To demonstrate the fidelity of this representation, we reconstruct beams at specific frequencies by applying the transformation parameters to the Zernike modes
\begin{equation}
    M_{f}(l, m) = \sum_{k=1}^{N_{\text{modes}}} c_k Z_k(s_f (l - \Delta l_{0,f}), s_f (m - \Delta m_{0,f})),
    \label{Eq: Zernike_modes}
\end{equation}
where $c_k$ are the Zernike coefficients and $(s_f, \Delta l_{0,f}, \Delta m_{0,f})$ 
are the frequency-dependent transformation parameters. Since Zernike polynomials are continuous analytic functions, this representation enables beam reconstruction at arbitrary spatial resolution, making it directly applicable to operational imaging and calibration pipelines.

\begin{figure*}
    \centering
    \includegraphics[width=\linewidth]{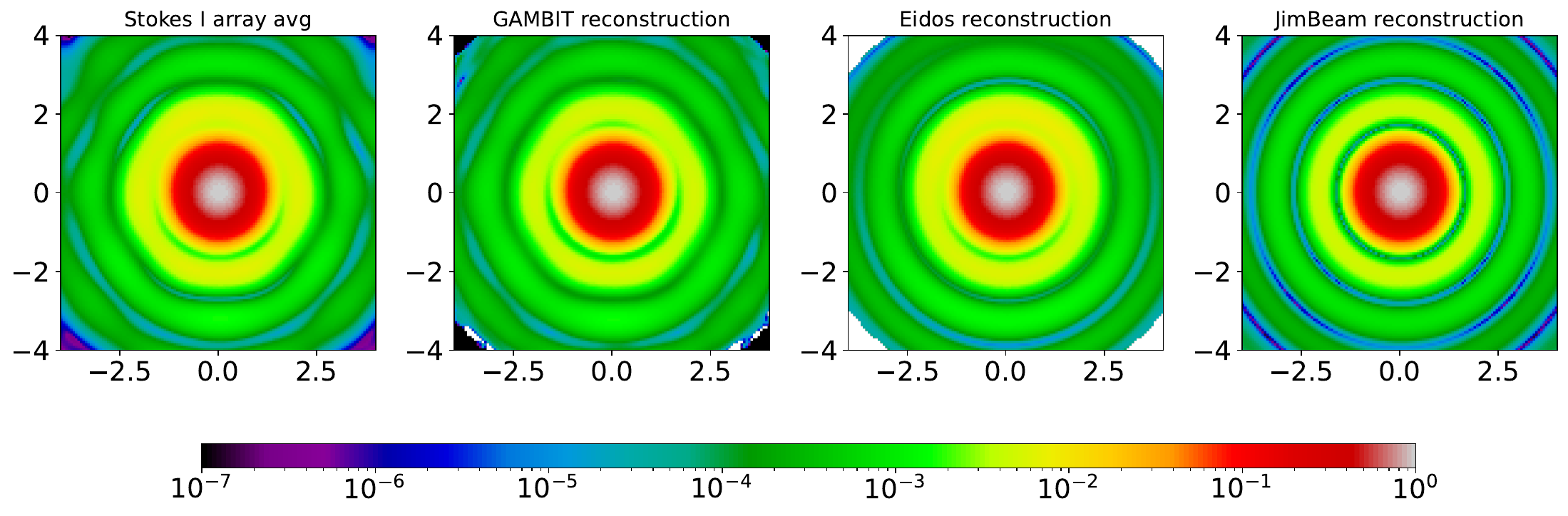}
    \caption{Comparison of beam reconstructions at 1107\,MHz. From left to right: Holographic Stokes~$I$ array average, smoothed GAMBIT reconstruction, Eidos model, and JimBeam model. All panels share a common logarithmic colour scale spanning seven decades. GAMBIT closely reproduces the sidelobe structure and ring morphology of the measured beam.}
    \label{fig:Zernike_recon_Gambit_Eidos_comparison}
\end{figure*}

Fig.~\ref{fig:Zernike_recon_Gambit_Eidos_comparison} shows the reconstructed beam at 1107\,MHz for the holographic array average, and the smoothed GAMBIT, Eidos, and JimBeam models. The GAMBIT model accurately reproduces the detailed sidelobe structure, including the faint hexagonal pattern, without imposing circular symmetry assumptions common to other models.

{}

\subsection{Computational efficiency}

A critical practical advantage of our generative Bayesian model is its efficient data compression. The holographic Stokes~$I$ measurements for a single antenna across the L band comprise approximately 67\,MB of data -- 1024 frequency channels in total, of which 900 are used in our analysis, each containing a full 2D spatial beam map. Our Bayesian framework compresses the resulting $\approx$60\,MB dataset into just 150\,kB: a single base beam pattern plus three sets of smooth, frequency-dependent transformation parameters (scale factors, $l_0$-shifts, and
$m_0$-shifts). This represents a compression ratio of 400:1, achieved, while preserving the essential physical structure of the beam and maintaining a reduced $\chi^2/\mathrm{dof} \sim 1$ across the band.

This data reduction is not lossy compression in the traditional sense—it provides some insight into the underlying physics. The model reveals that despite the apparent complexity of 900 independent beam patterns, the frequency variations are governed by smooth geometric transformations applied to a single base structure. The 150 kB model represents the physically meaningful degrees of freedom, while discarding measurement noise and redundant information.

A second key advantage of our approach is its computational efficiency at high spatial resolutions. In traditional beam modelling, increasing the resolution requires recomputing the full model at every frequency, which becomes prohibitively expensive for wide-bandwidth observations. In contrast, our framework requires building the base beam only once at the target resolution; reconstruction at any frequency is then reduced to applying the inferred scale and shift transformations—an effectively instantaneous operation. As a concrete benchmark, generating a high-resolution $8192 \times 8192$ (8k $\times$ 8k) base beam takes less than one minute on a 32 CPU-core, 62.5 GB RAM machine. This makes the framework directly applicable to high-resolution imaging and calibration pipelines, where competing models typically require substantially longer reconstruction times.

\section{Conclusion}

\label{sec:conclusion}
In this work, we have presented a Bayesian approach to modelling the primary beam of the MeerKAT antenna using the NIFTy framework. Our GAMBIT model employs Gaussian process priors to represent the beam as a continuous spatial field and incorporates frequency-dependent geometric transformations—specifically a scale factor and translational shifts—to capture the spectral variations in the beam. By fitting this generative model to holographic Stokes~I measurements across L band, we infer a posterior mean beam shape together with frequency-dependent scaling and centroid shift parameters that accurately reproduce the observed beam behaviour across the band.

We evaluated the performance of our Bayesian beam model against pervasive analytical and empirical models. The comparison was performed using the mean square deviation between model predictions and measured holographic beam maps. Our results demonstrate that the GAMBIT model consistently achieves the lowest deviation from the measured beam across the L band, outperforming all other models considered.

The subtle illumination of the outer parts of the MeerKAT main reflector panels, causing the observed hexagonal-like structure to sidelobe rings, is also expected in other modern radio telescopes, including SKA-Mid and the ngVLA. GAMBIT is able to capture these features directly from the holographic measurements.

Beyond accuracy, the model achieves exceptional computational efficiency on two fronts. First, it compresses 60~MB of holographic measurements into just 150~kB per antenna—approximately 400:1 compression—while maintaining full statistical consistency with the data. Second, it enables rapid high-resolution reconstruction: building the base beam at high resolution, for example 8k $\times$ 8k takes under one minute on a standard 32-core workstation, after which beam evaluation at any frequency reduces to applying the inferred scale and shift transformations, which are effectively instantaneous operations.

This makes the approach particularly well suited for incorporation into direction-dependent calibration schemes and primary beam correction workflows in radio interferometric imaging. Moreover, the framework is not limited to MeerKAT and can be readily adapted to other radio telescopes, offering a general pathway towards improved calibration precision and enhanced scientific output for current and future radio surveys.

In forthcoming publications, we shall extend GAMBIT to full Stokes
polarimetry and apply it to all 64 MeerKAT antennas across multiple frequency bands, providing a comprehensive, array-wide beam model suite for the community. The compact representation developed here—comprising a single base beam pattern and frequency-dependent transformations—enables rapid beam evaluation at arbitrary frequencies through simple geometric operations. This computational efficiency makes the models directly applicable to real-time calibration and imaging pipelines, where beam corrections can be computed on-the-fly during data processing without significant computational overhead. Such a resource will provide the radio astronomy community with accurate, readily deployable primary beam models for MeerKAT observations, facilitating improved calibration accuracy and enhanced image fidelity in science-ready data products.

\section{Data availability}

The beam model presented in this work can be generated using the accompanying GAMBIT 
Python package (\url{https://github.com/MortezaPasha/GAMBIT-beam-model}) developed as part of this work. The package implements the inferred base beam, frequency-dependent geometric transformations, reconstructed beam at each frequency and at arbitrary resolution, and Zernike-based decompositions described in the text. The code and instructions for reproducing the results will be made publicly available upon publication.

\begin{acknowledgements}
 We acknowledge financial support from the German Ministry  of  Research,  Technology  and Space  (BMFTR)  as  part  of  the  joint project ErUM-IFT (grant numbers: 05D23PB1, 05D23E01, 05D23WE2) and D-MeerKAT-III (grant number: 05A23PBA). AB gratefully acknowledges the financial support provided by the German Federal Ministry of Research, Technology and Space (BMFTR) in the framework of the Knowledge creates perspectives for the region!, for the project StStG – DZA – Aufbauphase: Deutsches Zentrum f\"ur Astrophysik, Gro{\ss}forschungszentrum in der s\"achsischen Lausitz: Aufbauphase 2026, grant number 03WSP1746.
\end{acknowledgements}

\bibliographystyle{aa}
\bibliography{aa}

\begin{appendix}
\section{Multi-antenna shift parameter comparison} \label{App:ShiftParameterComparison}

\begin{figure*}
    \centering
    \includegraphics[width=\textwidth]{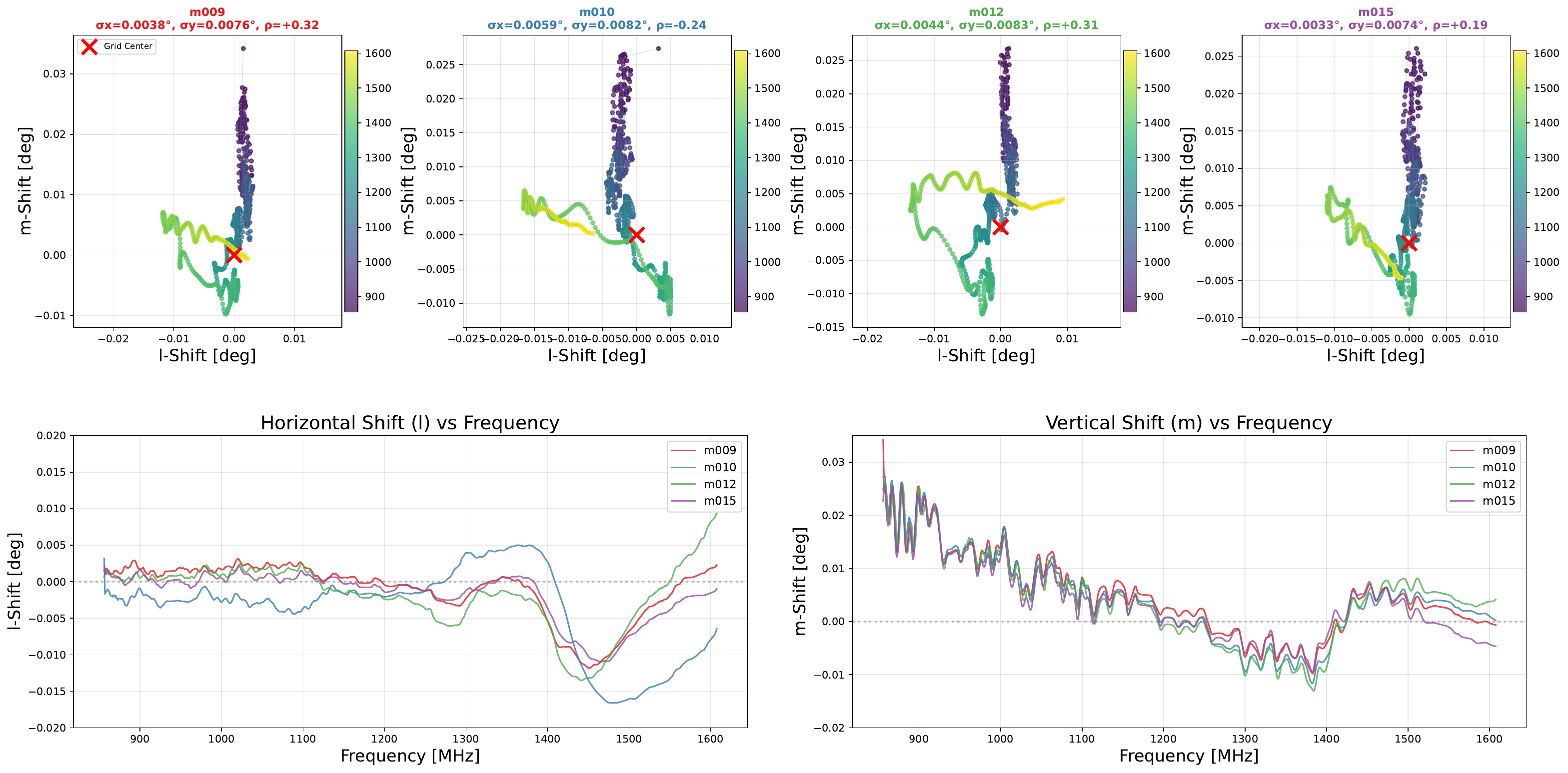}
    \caption{Centre-of-mass analysis of the beam centroid motion for four MeerKAT antennas (m009, m010, m012, m015). \textit{Top row:} Two-dimensional centroid trajectories in the $(l,\,m)$ plane, colour-coded by frequency (900--1600\,MHz); the red cross marks the nominal grid centre. \textit{Bottom:} Horizontal ($l$, left) and vertical ($m$, right) centroid shifts versus frequency.}
    \label{fig:CenterOfMass}
\end{figure*}

To assess the generalisability of the frequency-dependent shift behaviour across the MeerKAT array, we apply a centre-of-mass analysis to the holographic beam measurements from four antennas: m009, m010, m012, and m015. This analysis provides an independent, model-free characterisation of the beam centroid motion as a function of frequency.

The centre-of-mass shift $(\Delta l_{\rm COM}, \Delta m_{\rm COM})$ at each frequency is computed directly from the measured beam intensity distribution $I_{ij}(f)$ at pixel coordinates $(l_{ij}, m_{ij})$ as:
\begin{equation}
\Delta l_{\rm COM}(f) = \frac{\sum_{i,j} l_{ij} \, I_{ij}(f)}{\sum_{i,j} I_{ij}(f)} - l_{\rm centre},
\end{equation}

\begin{equation}
    \Delta m_{\rm COM}(f) = \frac{\sum_{i,j} m_{ij} \, I_{ij}(f)}{\sum_{i,j} I_{ij}(f)} - m_{\rm centre},
\end{equation}
where the sums run over all spatial pixels in the holographic map at frequency $f$, and $(l_{\rm centre}, m_{\rm centre})$ is the reference grid centre position (marked with a red cross in the top row of Fig.~\ref{fig:CenterOfMass}).

Fig.~\ref{fig:CenterOfMass} presents a comprehensive comparison of the centre-of-mass trajectories across all four antennas. The top row shows the two-dimensional shift trajectories in the $l$--$m$ plane, colour-coded by frequency (MHz). Each trajectory reveals a systematic variation from the upper portion of the parameter space (low frequencies, $\sim$900~MHz, shown in purple ) to the lower portion (high frequencies, $\sim$1600~MHz, shown in green-yellow), following a curved path that qualitatively resembles chromatic aberration in optical systems. However, in this radio-frequency context, the effect is more likely attributable to frequency-dependent propagation modes within the waveguide and feed horn system.

The bottom row decomposes the centre-of-mass shifts into horizontal (left panel) and vertical (right panel) components as a function of frequency. The vertical shift component exhibits striking consistency across all four antennas: a monotonic decrease from $\sim$0.03$\degr$ at 900~MHz to approximately $-$0.01$\degr$ at 1600~MHz, with only minor antenna-to-antenna variations. This universal trend suggests that the dominant vertical shift is driven by a systematic, array-wide effect---most plausibly related to the common feed horn and waveguide design shared by all MeerKAT antennas.

In contrast, the horizontal shift component is significantly smaller in magnitude (typically $<$0.01$\degr$) and displays greater antenna-to-antenna variability. Notably, antenna m010 exhibits a distinct frequency dependence compared to the other three antennas, particularly in the 1300--1600~MHz range, where it shows a pronounced negative excursion. This suggests that while the vertical shift is dominated by common systematic effects, the horizontal shift is more sensitive to antenna-specific asymmetries, such as variations in reflector surface quality, feed positioning tolerances, or structural alignment.

The statistical summaries provided in the top row quantify the mean centre-of-mass shifts and their correlation: $\sigma_l$ and $\sigma_m$ represent the standard deviations of the $x$- and $y$-shifts across frequency, respectively, while $\rho$ denotes the Pearson correlation coefficient between $\Delta l_{\rm COM}$ and $\Delta m_{\rm COM}$. The correlations are uniformly small, ranging from $\rho = -0.24$ (m010) to $\rho = +0.32$ (m009, m012), indicating that the horizontal and vertical shifts are approximately independent and can be modelled separately without significant loss of information.

These centre-of-mass results demonstrate that the frequency-dependent shift behaviour observed for antennas m009, m010, m012, and m015 is representative of the broader MeerKAT array, with the vertical component showing strong inter-antenna consistency and the horizontal component exhibiting smaller, antenna-specific variations. This independent analysis supports the physical interpretation of the geometric transformation parameters inferred in the main Bayesian model and validates the generalisability of our framework.

\section{Noise estimation across frequency}
\label{App:NoiseAnalysis}

\begin{figure*}
    \centering
    \includegraphics[width=\textwidth]{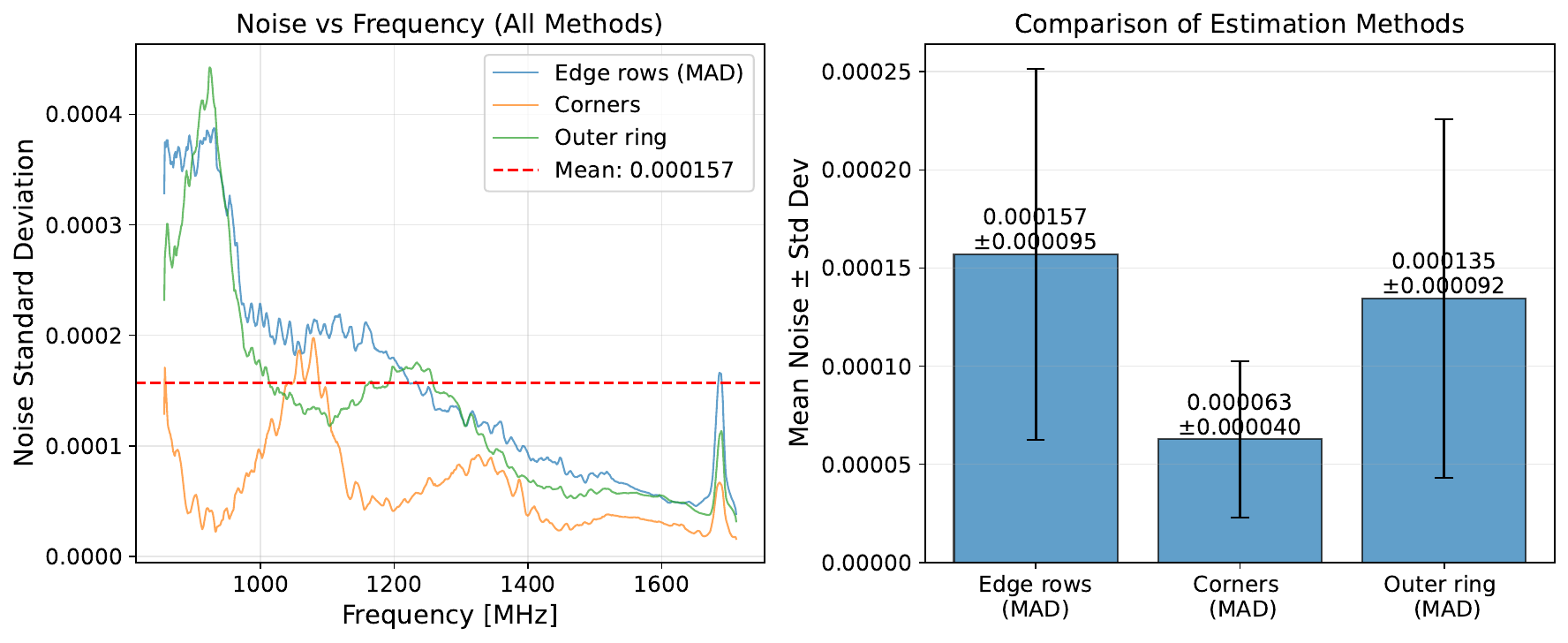}
    \caption{Noise characterisation of the holographic beam maps using three MAD based spatial estimators: edge rows (blue), corner regions (orange), and the outer ring (green). \textit{Left:} Noise standard deviation versus frequency; the dashed red line marks the band-averaged edge-row mean ($1.57\times10^{-4}$). \textit{Right:} Band-averaged mean and standard deviation per estimator. Corner regions yield the lowest ($6.3\times10^{-5}\pm4.0\times10^{-5}$).}
    \label{fig:Noise analysis}
\end{figure*}

Fig.~\ref{fig:Noise analysis} presents the estimated noise level as a function of frequency across the L band, obtained using three different spatial regions of the holographic beam maps: the image edges (outer rows), the outer ring, and the corner regions. The noise is quantified using MAD estimator, which provides a robust measure in the presence of potential signal contamination.

The left panel shows the frequency-dependent noise estimates for all three methods. While all approaches exhibit broadly similar trends, the edge and outer-ring estimates display larger scatter and systematically higher values at certain frequencies. This is expected, as these regions are more likely to contain residual beam signal, particularly from sidelobes, leading to an overestimation of the noise level.

In contrast, the corner-based estimates are consistently lower and exhibit reduced variance across frequency, indicating minimal contamination from the beam signal. This suggests that the corners provide a more reliable proxy for the intrinsic noise level of the measurements.

The right panel summarises the mean noise level and its standard deviation for each method. The corner-based approach yields the lowest mean noise estimate, $\sigma \approx 6.3 \times 10^{-5}$, while the edge and outer-ring methods produce higher mean values of $\sim 1.6 \times 10^{-4}$ and $\sim 1.3 \times 10^{-4}$, respectively, with larger associated uncertainties. Despite these differences, the estimates remain consistent within their respective uncertainties, particularly in the mid-band region.

Given the potential for residual signal contamination and the lack of a clear physical model for frequency-dependent noise variations, we adopt a single representative mean noise level across all frequencies in the main analysis. This choice provides a stable and conservative approximation of the noise properties for the Bayesian inference.

\section{Dominant Zernike polynomials of base beam and residuals} \label{app:ZernikeBaseBeam}

The inferred base beam $B(l, m)$ is decomposed into Zernike polynomial modes to provide a compact, physically interpretable representation. Zernike polynomials form a complete orthogonal basis on the unit disc and correspond to classical optical aberrations, making them particularly suited for antenna beam analysis.

\begin{figure*}
    \centering
    \includegraphics[width=\textwidth]{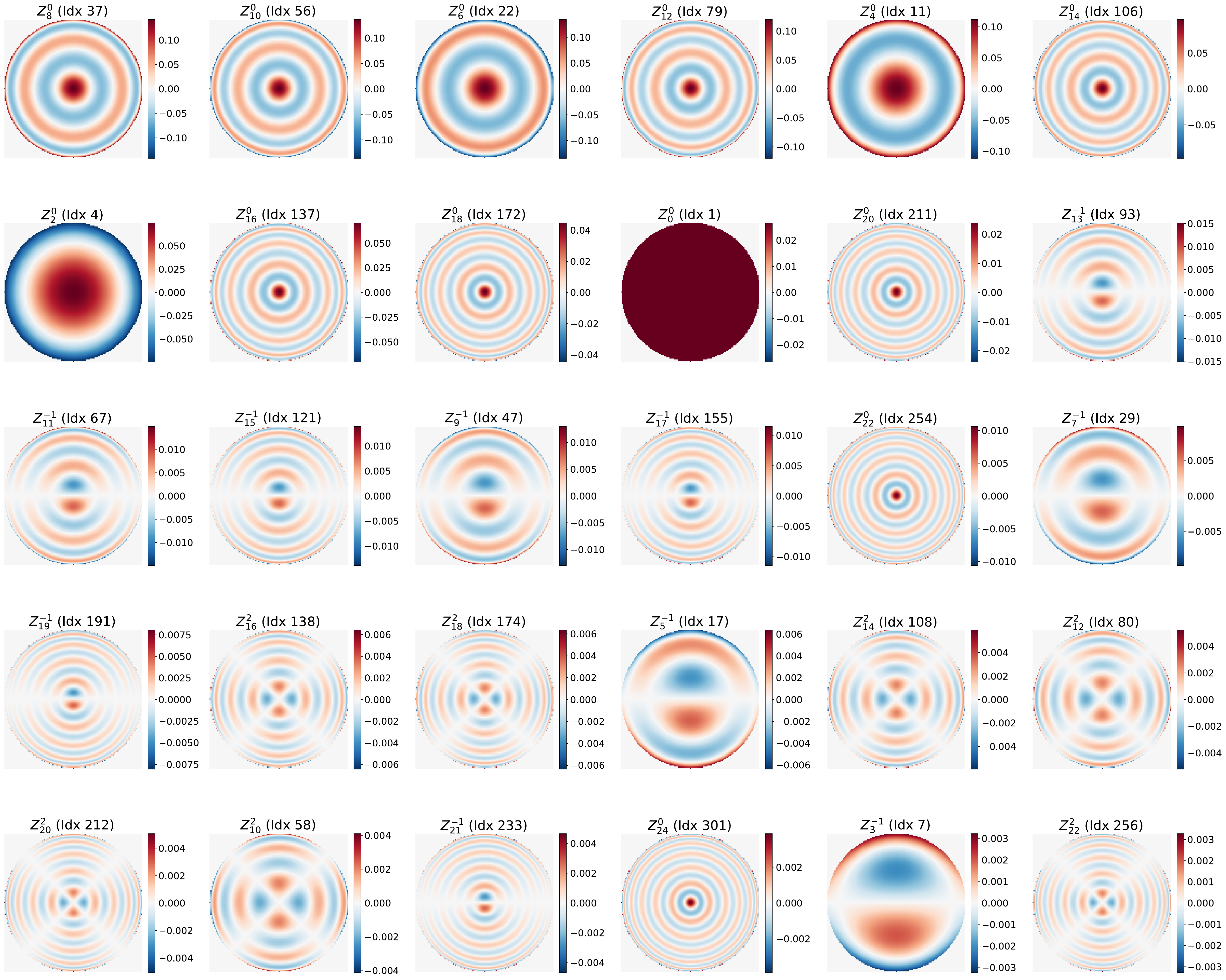}
    \caption{ Spatial visualisation of the top 30 Zernike modes ($Z_n^m$) decomposing the base beam, ranked by coefficient magnitude. Each panel shows the two-dimensional pattern of a Zernike polynomial with its coefficient value indicated by the colour bar. Radially symmetric modes ($m=0$) dominate with coefficients of $0.02-0.10$, while asymmetric modes ($m=\pm1$ coma, $m=\pm2$ astigmatism) contribute at the $0.003-0.010$ level. The clear hierarchy in coefficient magnitudes reflects the physical dominance of radial beam structure over directional aberrations.
}
    \label{fig:zernike_top30_coefficients}
\end{figure*}

Fig.~\ref{fig:zernike_top30_coefficients} displays the spatial structure of the 30 dominant Zernike modes for the MeerKAT array-average base beam. The decomposition reveals that the beam structure is dominated by low-order azimuthally symmetric modes ($Z_n^0$), which encode defocus and spherical aberration. Asymmetric modes with $m \neq 0$ contribute at lower amplitudes, representing coma, astigmatism, and higher-order aberrations.

\begin{figure*}
    \centering
    \includegraphics[width=\textwidth]{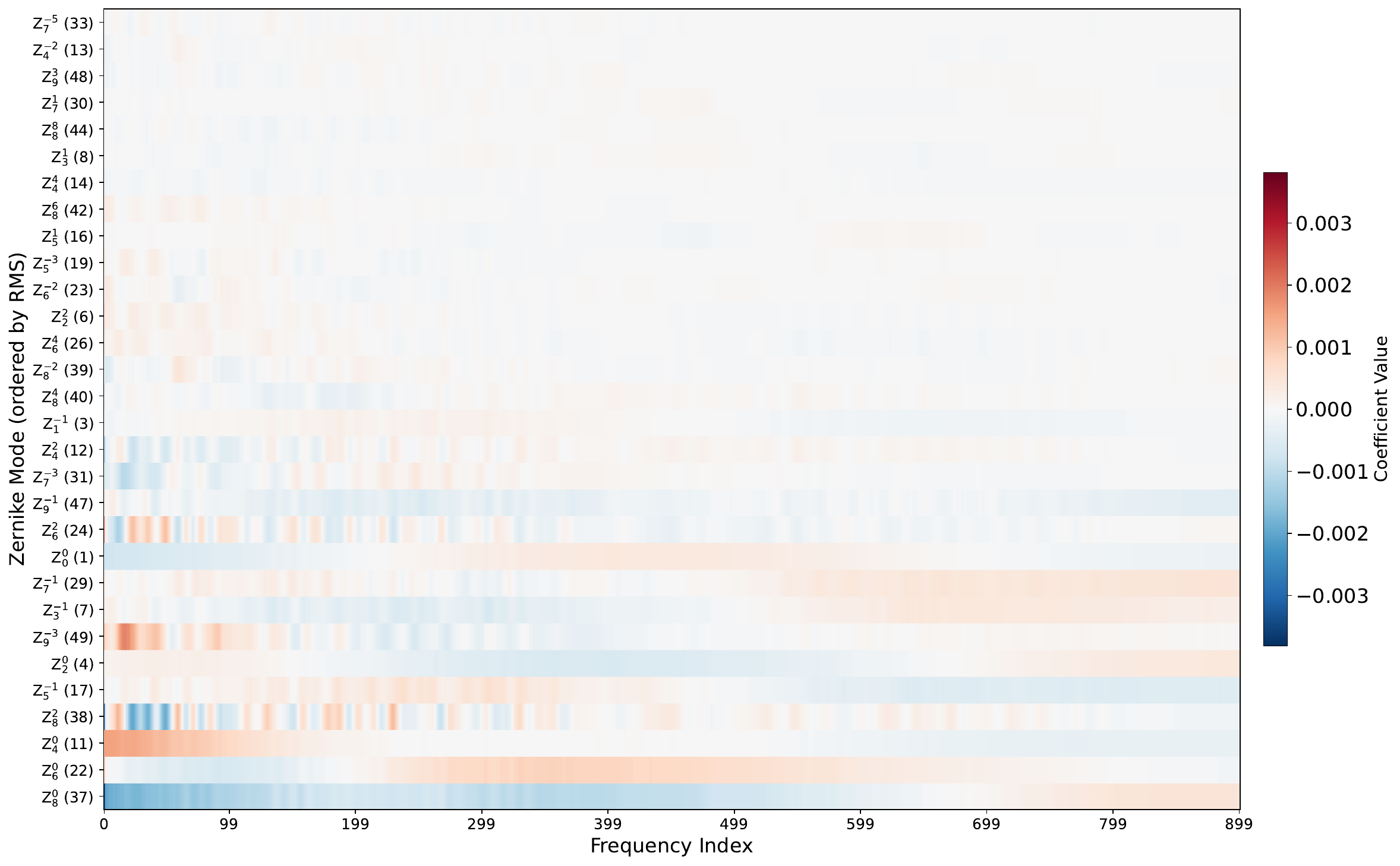}
    \caption{Frequency-dependent Zernike decomposition of beam model residuals (measured beam minus reconstructed model). The heatmap displays coefficient values for 30 Zernike modes (vertical axis, ordered by root-mean-square coefficient magnitude) across 900 frequency channels spanning the L band (horizontal axis). The colour scale indicates residual coefficient values, with white representing zero residual. Several modes exhibit systematic frequency-dependent structure with coefficient variations of order $\pm0.003$, revealing subtle beam shape variations not captured by the geometric transformation model. }
    \label{fig:zernike_residual_heatmap}
\end{figure*}

Fig.~\ref{fig:zernike_residual_heatmap} shows the Zernike decomposition of the residuals (data minus our model) across frequency for the top 30 modes. Several modes exhibit coherent, smooth variations with frequency, indicating genuine chromatic effects not captured by global geometric transformations alone. These patterns likely arise from frequency-dependent feed illumination changes or subtle optical path variations.

Although our hierarchical model employs frequency-independent Zernike coefficients by design—with spectral structure captured through scaling and shift transformations—an analysis of the residuals (data minus model) reveals systematic frequency-dependent structure. 
Fig.~\ref{fig:zernike_residual_heatmap} shows the Zernike decomposition of these 
residuals across frequency for the top 30 modes.

The current frequency-independent Zernike representation, combined with scaling-shift transformations, successfully captures the dominant beam variations (as demonstrated by the low MSD values in the main text), while these residual patterns provide diagnostic information about higher-order chromatic aberrations in the MeerKAT optical system.

\end{appendix}

\end{document}